\documentclass[prd, twocolumn, amsmath,amssymb,aps, reprint, preprintnumbers, nofootinbib, superscriptaddress]
{revtex4-2}
\usepackage{xcolor}
\usepackage{amsmath}
\usepackage{physics}
\usepackage{amssymb}
\usepackage{graphicx}
\usepackage{svg}
\usepackage{subcaption}
\usepackage{array}
\usepackage{tikz}
\usetikzlibrary{tikzmark,calc,shapes.geometric,arrows.meta}
\usepackage{diagbox}
\usepackage{multirow}
\usepackage[toc,page]{appendix}
\usepackage{url}
\usepackage{ragged2e}
\pdfoutput=1
\usepackage{amsfonts}
\usepackage{mathrsfs}
\usepackage{graphicx}
\usepackage{color}
\usepackage{bm}
\usepackage{blindtext}
\usepackage{wasysym}
\usepackage{hyperref}
\hypersetup{colorlinks=true,allcolors=blue}
\usepackage[normalem]{ulem}

\begin{document}

\begin{abstract}In the Standard Model Effective Field Theory (SMEFT), the $SU(2)_L\times U(1)_Y$ symmetry of the Standard Model is linearly realized. However, it is possible that more general effective field theories, such as the Higgs Effective Field Theory (HEFT) where this symmetry is realized non-linearly, are needed to describe the data. Identifying physics beyond SMEFT could shed light on the nature of Higgs and the realization of the electroweak symmetry. We explore the possibility of such an identification by studying the effects of scalar and vector new-physics operators on the angular distribution of $\Lambda_b\rightarrow \Lambda_c(\rightarrow\Lambda\pi)\tau\bar\nu_\tau$. This decay is sensitive to the 6-dimensional effective operator $O_V^{LR}\equiv(\bar{\tau}\gamma^\mu P_L\nu_\tau)(\bar{c}\gamma_\mu P_R b)$, which is present in HEFT but suppressed in SMEFT. We identify the angular observables that can have significant contributions from $O_V^{LR}$, and hence would be useful for probing not only BSM physics but also physics beyond SMEFT. We further find that constraining the branching ratio of $B_c\rightarrow \tau \bar \nu_\tau$ would be crucial for performing this task.
\end{abstract}


\title{\boldmath Identifying physics beyond SMEFT in the \\ angular distribution of $\Lambda_b\rightarrow \Lambda_c(\rightarrow\Lambda\pi)\tau\bar\nu_\tau$ decay}

\author{Siddhartha Karmakar}
\email{siddhartha@theory.tifr.res.in (ORCID: 0009-0003-0609-9689)}
\affiliation{Tata Institute of Fundamental Research, Homi Bhabha Road, Colaba, Mumbai 400005, India}
\author{Susobhan Chattopadhyay}
\email{susobhan.chattopadhyay@tifr.res.in (ORCID: 0009-0000-2346-2273)}
\affiliation{Tata Institute of Fundamental Research, Homi Bhabha Road, Colaba, Mumbai 400005, India}
\author{Amol Dighe} 
\email{amol@theory.tifr.res.in (ORCID: 0000-0001-6639-0951)}
\affiliation{Tata Institute of Fundamental Research, Homi Bhabha Road, Colaba, Mumbai 400005, India}

\maketitle

\section{Introduction}
The Standard Model (SM) of particle physics describes the properties of known matter and forces to a great accuracy. However, there are reasons to believe that the SM is not complete. Certain phenomena such as the baryon asymmetry in the universe, neutrino oscillations, and the existence of dark matter are not explained in the SM. This indicates that there exists new physics (NP) beyond SM. One way this NP can be probed in direct searches is by producing new particles in high energy particle colliders. In a complementary approach, the effects of NP can be identified indirectly with the help of precision measurements in relatively low-energy processes.

While the effect of NP can be searched for in specific models, the effective field theory (EFT) techniques provide a model independent and common framework to perform more general analyzes, whose results can be translated into broad classes of scenarios beyond the Standard Model (BSM). In the field of flavor physics, EFT methods are employed to analyze the effect of NP with the help of precision measurements.

In the modern view, SM is considered as the leading part of an EFT that respects the SM symmetry group $SU(3)_C\times SU(2)_L\times U(1)_Y$. In this view, the SM is expected to be valid up to an NP scale $\Lambda$, above which additional dynamical degrees of freedom appear. The effects of NP can be encoded in higher-dimensional operators which are suppressed by powers of the NP scale $\Lambda$:
\begin{align}
	\mathcal{L}&= \mathcal{L}_{SM}+\frac{1}{\Lambda}C^{(5)}O^{(5)}+\frac{1}{\Lambda^2}\sum_{i}C_i^{(6)}O_i^{(6)}+\mathcal{O}\left(\frac{1}{\Lambda^3}\right).\label{SMEFTexpansion}
\end{align}
The above equation describes the Standard Model Effective Field Theory (SMEFT)~\cite{Buchmuller:1985jz, Grzadkowski:2010es, Jenkins:2013zja, Isidori:2023pyp} when all the operators $O_i^{(d)}$ are composed of only SM fields and respect the SM symmetry group. Here $d$ denotes the dimensionality of the operator $O_i^{(d)}$.  The NP contributions are parameterized in terms of the Wilson coefficients $C_i^{(d)}$. 

In SM, the Higgs boson $(h)$ and the three Goldstone bosons $(\phi_1,~\phi_2,~\phi_3)$ are parts of a single $SU(2)_L$ doublet field $H$. This embedding is referred to as a linear realization of the electroweak (EW) symmetry~\cite{Feruglio:1992wf}. This feature is maintained in SMEFT. At present, the measurements of production and decay channels of the Higgs boson at the LHC are consistent with the SM Higgs mechanism of EW symmetry breaking, and hence with SMEFT~\cite{LHCHiggsCrossSectionWorkingGroup:2016ypw}. However, in BSM scenarios, the four fields $(h,~\phi_1,~\phi_2,~\phi_3)$ need not be embedded into a single doublet $H$.  Such scenarios cannot be described in SMEFT, and searching for these scenarios is equivalent to probing the mechanism of EW symmetry breaking.

Possible deviations of couplings of the Higgs boson from the SM prediction would indicate the need for a more general extension of SM, even beyond SMEFT. Note that the measured values of the Higgs boson couplings to gauge bosons and top quark are compatible with SM to about $\mathcal{O}(10\%)$~\cite{DiMicco:2019ngk, Grober:2843280,CMS:2022dwd}. However, for the Higgs boson couplings to other quarks and leptons, or for the triple-Higgs boson coupling, larger deviations are possible~\cite{ DiMicco:2019ngk,Grober:2843280, CMS:2022dwd}. For example, in the Higgs Effective Field Theory (HEFT)~\cite{ Alonso:2012px, Buchalla:2013rka, Pich:2016lew}, one can have modified interactions of the Higgs boson while keeping its gauge interactions unchanged. Here the Higgs sector consists of a singlet $\hat h$ boson that is invariant under the EW symmetry and three Goldstone fields $(\hat\phi_1,~\hat\phi_2,~\hat\phi_3)$ that transform non-linearly under this symmetry~\cite{Feruglio:1992wf}. Thus, the $SU(2)_L\times U(1)_Y$ symmetry is realized non-linearly in HEFT, and the manifest gauge symmetry in HEFT is only $SU(3)_C\times U(1)_{Q}$. Therefore, HEFT is a more general effective theory than SMEFT, i.e. $\textrm{HEFT}\supset \textrm{SMEFT}\supset\textrm{SM}$~\cite{Cohen:2020xca}.  In particular, many HEFT operators at a given dimension may not appear in SMEFT at the same dimension, though they may appear at higher dimensions and therefore would typically be suppressed.

Searches for beyond-SMEFT physics in ATLAS and CMS focus on the precise measurements of Higgs couplings with fermions and gauge bosons~\cite{DiMicco:2019ngk}. A complementary approach to extract evidence for non-linearly realized EW symmetry is via flavor physics, which offers indirect probes of heavy new physics that are complementary to direct collider searches. 
Many of the recently observed anomalies in $B$ meson decays, e.g. $R(D^{(*)})$~\cite{BaBar:2012obs,BaBar:2013mob,Belle:2015qfa,LHCb:2015gmp} and  $R(J/\psi)$~\cite{LHCb:2017vlu} that correspond to the charged-current transition $b\rightarrow c \tau \nu_\tau$, and $\mathcal{B}(B^+\rightarrow K^+\mu^+\mu^-)$~\cite{LHCb:2014cxe}, $\mathcal{B}(B^+\rightarrow K^+e^+e^-)$~\cite{LHCb:2022zom} and $P^\prime_5$~\cite{LHCb:2013ghj, Descotes-Genon:2012isb,Descotes-Genon:2013wba} that correspond to the neutral-current transitions $b\rightarrow s l l$,  indicate the possibility of BSM physics. 
While there are viable solutions to these anomalies within SMEFT, the question of whether these anomalies arise from physics beyond SMEFT is still open.

For flavor physics processes, the relevant energy scale is around the mass of the $b$ quark. At this 
energy scale, the heavier SM particles $(W^{\pm}, Z^0, h, t)$ are no longer the degrees of freedom and are integrated out. The resultant EFT is called the Low-energy Effective Field Theory (LEFT),\footnote{LEFT is sometimes referred to as weak effective field theory (WET or WEFT) in literature~\cite{Jenkins:2017jig,Aebischer:2017gaw,Aebischer:2017ugx, London:2021lfn}.}~\cite{Buchalla:1995vs} in which the effective Lagrangian is 
\begin{align}
	\mathcal{L}_{\textrm{LEFT}} &= \mathcal{L}_{\textrm{SM}} + \sum_{d\ge 5}\sum_{i}(G_F)^{\frac{d}{2}-2}\,C_i^{(d)}\mathcal{O}^{(d)}_i~.
\end{align}
Here, the expansion is in terms of the Fermi constant $G_F=(\sqrt 2 g_W^2)/(8 M_W^2)$ .The operators are written in terms of the dynamical fields below the EW symmetry-breaking scale. Thus, LEFT is an $SU(3)_C\times U(1)_Q$ invariant effective theory valid below the EW scale. 

The information about how the $SU(2)_L\times U(1)_{Y}$ symmetry is realized above the EW scale can be extracted by matching the LEFT operators to EFTs valid beyond the electroweak scale, such as SMEFT or HEFT. Most of the dimension-6 operators do not require any source of electroweak symmetry breaking beyond SM and these operators can be directly mapped onto six-dimensional SMEFT and HEFT operators~\cite{Aebischer:2015fzz, Jenkins:2017dyc, Jenkins:2017jig}. However, there are some dimension-6 LEFT operators which cannot be generated from dimension-6 SMEFT operators but can be generated from HEFT at this order~\cite{Alonso:2014csa, Cata:2015lta, Burgess:2021ylu}. There exist NP models where certain dimension-6 operators are generated but cannot be mapped to SMEFT at the leading order --- for example, the non-standard Higgs model with a strongly-coupled scalar~\cite{Cata:2015lta} and the model with a $W^\prime$ that couples to right-handed quarks and left-handed leptons~\cite{Jacob:1959at}. Further discussions on models requiring a framework beyond SMEFT can be found in~\cite{Banta:2021dek}, where a new class of BSM states called `Loryons' is proposed. Loryons are NP particles whose physical mass is dominated by a contribution from the vacuum expectation value of the Higgs boson. However, the effect of integrating out such fields cannot be incorporated into SMEFT and would require an EFT framework where the EW symmetry is non-linearly realized. The study of such operators will allow us to identify new physics beyond SMEFT.

One such LEFT operator is $O_V^{LR}\equiv (\bar \tau \gamma^\mu P_L \nu)(\bar c \gamma_\mu P_R b)$~\cite{Buchalla:2012qq,Cata:2015lta,Burgess:2021ylu}, which corresponds to the quark level transition $b\rightarrow c\tau\nu_\tau$. This operator can be generated at the leading order from the HEFT operator  $\mathcal{O}_{FY11}=(\bar{l}UP_-r)(\bar{r}P_+U^\dagger l)$~\cite{Buchalla:2012qq}. There is no SMEFT operator that can directly yield such a flavor non-universal $O_V^{LR}$ at the low scale. The SMEFT dimension-6 operator $O_{Hud}=(H^\dagger  i  \overleftrightarrow{D}_\mu H)(\bar{u}\gamma^\mu d)$ induces an anomalous $b$-$c$-$W$ coupling and contributes to  $O_V^{LR}$~\cite{Cirigliano:2009wk, Aebischer:2015fzz}, but here the $W$ coupling to the leptons is flavor-universal and would not contribute to the $b\rightarrow c\tau\nu_\tau$ anomalies. To generate the operator  $O_V^{LR}$ only in the $\tau$ sector, we need the dimension-8 SMEFT operator $O_{l^2udH^2}=(\bar{l}d H)(\tilde{H}^\dagger \bar{u}l)$, whose coefficient would be suppressed by an extra factor of $v^2/\Lambda^2$.  For this reason, $O_V^{LR}$ is neglected in many EFT analyzes based on SMEFT (e.g.~\cite{Blanke:2018yud,Blanke:2019qrx,Boer:2019zmp, Hu:2018veh, Becirevic:2022bev}). However, in the context of mesons, fits to data on $R(D^{(*)})$,  $R(J/\psi)$ and $\mathcal{B}(B_c\rightarrow\tau^+\nu)$ have indicated that significant contributions from this operator are possible~\cite{Burgess:2021ylu, Ray:2019gkv, Huang:2022fke}.

A potential baryonic process involving $b\rightarrow c\tau \nu_\tau$, which could also offer avenues of probing $O_V^{LR}$, is the decay $\Lambda_b\rightarrow \Lambda_c\tau\bar\nu_\tau$. The baryons $\Lambda_b$ and $\Lambda_c$ are spin-half particles. As a result, all the vector and the scalar effective operators affect $\Lambda_b\to \Lambda_c \tau \bar \nu_\tau$ decay. In contrast, for the mesonic mode $B \to D$, the axial vector and the pseudoscalar operators do not contribute at the leading order, and for $B \to D^*$, the scalar operator does not contribute at the leading order ~\cite{Hu:2020axt}. Furthermore, the form factors in the baryonic mode are different from those in the mesonic modes. Thus, the baryonic mode offers observables complementary to those in the mesonic modes to study the new physics effects.  The large production cross-section of $\Lambda_b$ at LHC and the well-known form factors for $\Lambda_b\to \Lambda_c$ make this baryonic decay mode a good candidate to complement its mesonic counterpart. Recently, the first observation of this process was reported at LHCb with a significance of $6.1\sigma$~\cite{LHCb:2022piu}. The angular distribution of final state particles in $\Lambda_b\rightarrow \Lambda_c\tau\bar\nu_\tau$ process~\cite{Shivashankara:2015cta, Dutta:2015ueb,Detmold:2015aaa, Li:2016pdv,DiSalvo:2018ngq, Hu:2018veh, Ray:2018hrx, Penalva:2019rgt} and further in  $\Lambda_b\rightarrow \Lambda_c(\rightarrow\Lambda\pi)\tau\bar\nu_\tau$ process~\cite{  Gutsche:2015rrt, Boer:2019zmp,Ferrillo:2019owd,Mu:2019bin, Hu:2020axt,Becirevic:2022bev} offers multiple observables and ratios of observables where some of the systematic and hadronic uncertainties would cancel.  We identify the coefficients of terms in this angular distribution, and their ratios, which would be sensitive to $O_V^{LR}$. We also point out those observables that can show distinct signatures of the $O_V^{LR}$ contribution. 

In section~\ref{sec:angdist}, we present the angular distribution of $\Lambda_b\rightarrow \Lambda_c(\rightarrow\Lambda\pi)\tau\bar\nu_\tau$ decay in terms of the three physical angles involved therein. We calculate angular observables in terms of helicity amplitudes.  The efficiencies of these angular observables for identifying the effects of $O_V^{LR}$ are analyzed in section~\ref{sec:results}. We present our concluding remarks in section~\ref{sec:conclusion}. Appendix \ref{app:moments} provides the weighting functions that can be used to extract the angular  observables using the method of angular moments. Appendix \ref{app:mapping} maps our expressions for angular distribution with earlier literature in order to clarify the notation. In Appendix \ref{app: bin-wise}, we show a comparison between the effectiveness of the angular observables in different $q^2$ bins for identifying the effects of $O_V^{LR}$ with two possible upper bounds on $\mathcal{B}(B_c\rightarrow \tau \bar\nu_\tau)$.

\section{Angular distribution of $\Lambda_b\rightarrow\Lambda_c(\Lambda\pi)\tau\bar{\nu}_\tau$ decay}\label{sec:angdist}

In the language of LEFT, the effective Hamiltonian for the quark-level transition $b\rightarrow c \tau\bar {\nu}_\tau$ may be written as~\cite{Datta:2017aue,Boer:2019zmp}
\begin{align}
	\mathcal{H}_{\textrm{eff}} &= \frac{4\,G_FV_{cb}}{\sqrt{2}}\left[(1+g_L)\mathcal{O}_V^{LL}+g_R\mathcal{O}_V^{LR}\right.\nonumber\\
	&\left.~+g_S\mathcal{O}_S+g_P\mathcal{O}_P+g_T\mathcal{O}_T\right]\,.\label{eft}
\end{align}
Here, $G_F$ is the Fermi constant  and $V_{cb}$ is the Cabibbo-Kobayashi-Maskawa (CKM) matrix element. The effective operators are
\begin{align}
	\mathcal{O}_V^{LL}&=(\bar \tau \gamma_\mu P_L\nu_\tau)(\bar c \gamma^\mu P_L b),& \mathcal{O}_S &=\frac{1}{2}(\bar \tau P_L\nu_\tau)(\bar c b) ,\label{LEFTop1}\\
	\mathcal{O}_V^{LR}&=(\bar \tau \gamma_\mu P_L\nu_\tau)(\bar c \gamma^\mu P_R b) ,&  \mathcal{O}_P&=\frac{1}{2}(\bar \tau P_L\nu_\tau)(\bar c  \gamma_5 b),\nonumber\\
	\mathcal{O}_T&=(\bar \tau \sigma_{\mu\nu} P_L \nu_\tau)(\bar c \sigma^{\mu\nu}b)\,.\label{LEFTop2}
\end{align}
The corresponding NP Wilson coefficients are $g_L$, $g_R$, $g_S$, $g_P$ and $g_T$. In our analysis, we ignore the tensor operator for simplicity and only consider the vector and the scalar new physics operators. Among the operators in eqs.\,(\ref{LEFTop1}-\ref{LEFTop2}), only $O_V^{LR}$ cannot be mapped to a dimension-6 SMEFT operator. Our goal is to find observables that can distinguish the effect of this operator from SM and from other NP operators. A large non-zero value of $\mathcal{O}(0.1 \textrm{ -- } 1)$ for the coefficient $g_R$ would imply new physics beyond SMEFT.

\begin{figure}[h!]
 	\centering
 	\includegraphics[width=0.8\linewidth]{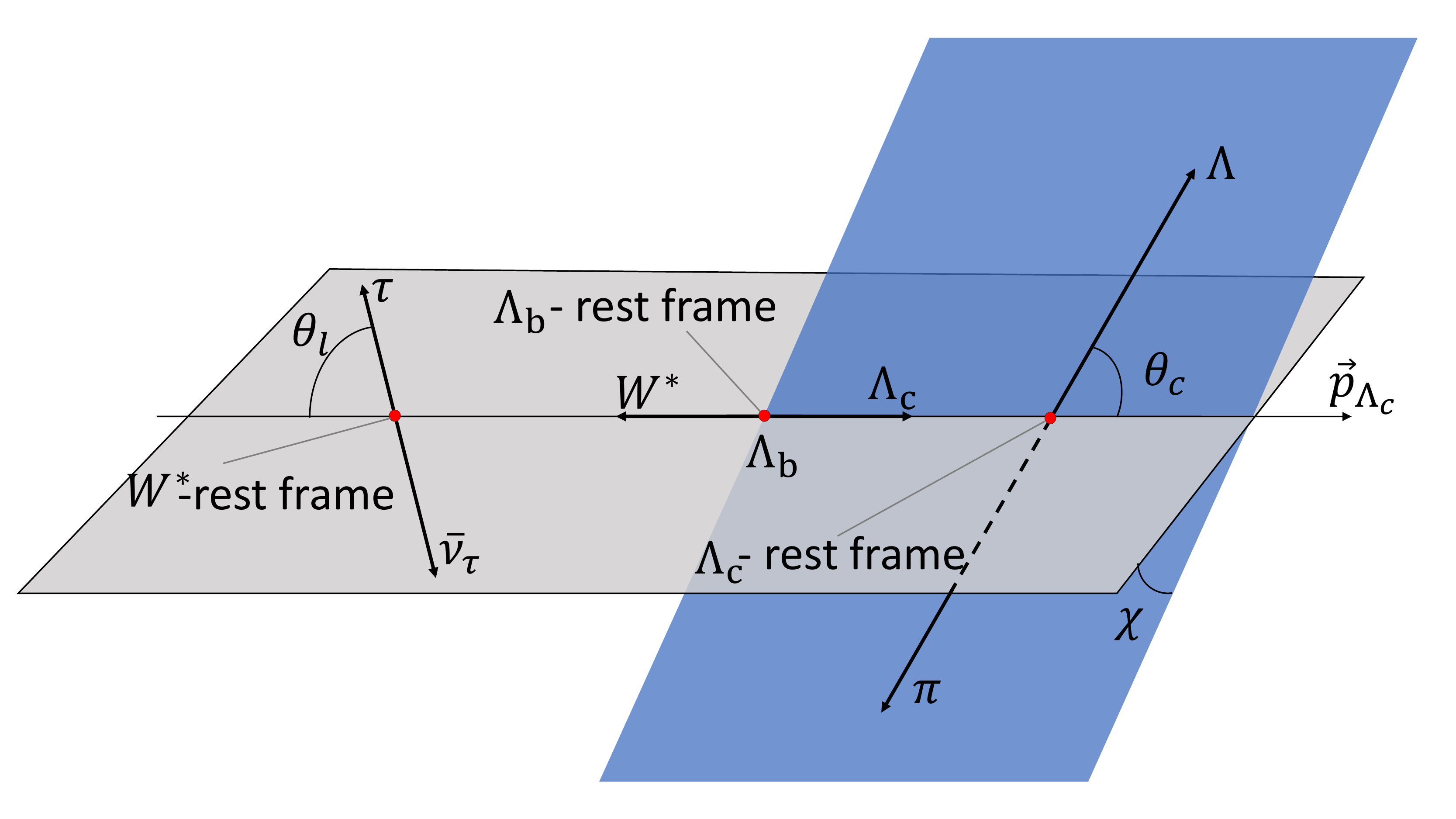}
 	\caption{\justifying Kinematics of $\Lambda_b\rightarrow \Lambda_c(\rightarrow \Lambda \pi)W^*(\rightarrow\tau\bar\nu_\tau)$ decay. The  angles are defined following~\cite{kutsckhe1996angular}.}
 	\label{angles} 
\end{figure}

The kinematics of the decay is shown in figure~\ref{angles}.
The spin components of the particles involved in the decay $\Lambda_b\rightarrow \Lambda_c(\rightarrow\Lambda\pi)W^*(\rightarrow\tau\,\bar\nu_\tau)$ may be represented as \\
\begin{center}\vspace{-.5cm}
	\begin{tikzpicture}[node distance=1cm]
		\node (Lb) {$\Lambda_b(s_1, m_1)$};
		\node (Lc) [right of=Lb, xshift=2cm] {$\Lambda_c(s_2,\lambda_2)~~$};
		\node (add1) [right of=Lc] {$+$};
		\node (W) [right of=add1, xshift=.2cm] {$~~W^*(s_3,\lambda_3)$};
		\node (Lpi) [below of=Lc, xshift=-3cm] {$\Lambda(\lambda_4)~+~\pi(\lambda_5)$};
		\node (taunu) [below of=W,xshift=-3cm, yshift=-1cm] {$\tau(\lambda_6)~+~\bar \nu_\tau(\lambda_7)~.$};
		\draw [->] (Lb) -- (Lc);
		\draw [->] (Lc) |- (Lpi);
		\draw [->] (W) |- (taunu);
	\end{tikzpicture}
\vspace{-.2cm}
\end{center}
Here, $s_1$ denotes the spin of $\Lambda_b$, and $m_1$ denotes its component  along an arbitrarily chosen  quantization axis. The spins and helicities of $\Lambda_c$ and $W^*$  in the $\Lambda_b$ rest frame are denoted by  $(s_2, \lambda_2)$ and $(s_3, \lambda_3)$, respectively. The helicities of  $(\Lambda,\pi)$ in the $\Lambda_c$ rest frame  are  $(\lambda_4,\lambda_5)$, while the helicities of $(\tau,\nu_\tau)$, in their center of mass frame, i.e. $W^*$ rest frame, are $(\lambda_6, \lambda_7)$. Clearly, $\lambda_5 = 0$ and $\lambda_7 = +1/2$.

\subsection{Definitions of the angles}

The angles in our analysis are defined as follows. In the rest frame of $\Lambda_b$, the $\Lambda_c$ hadron makes an angle $\theta_1$ with the spin quantization axis $\mathbf{z_1}$ of $\Lambda_b$. In this frame, the direction of $\Lambda_c$ momentum is taken to be the $\mathbf{z_2}$ axis, and the opposite direction, i.e. the direction of  $W^*$ momentum, is denoted as the $\mathbf{z_3}$ axis.  The x-axes $(\mathbf{x_1,x_2,x_3})$ are chosen arbitrarily in the planes orthogonal to $(\mathbf{z_1, z_2, z_3})$, keeping $\mathbf{x_2}=-\mathbf{x_3}$. This fixes the directions of the y-axes $(\mathbf{ y_1,y_2,y_3 })$ automatically. In the rest frame of $\Lambda_c$, the momentum of $\Lambda$ is denoted by the spherical polar coordinates $(\theta_2,\phi_2)$, while in the rest frame of $W^*$, the direction of $\tau$ is denoted by $(\theta_3, \phi_3)$.

The angular distribution of the decay $\Lambda_b\rightarrow\Lambda_c(\Lambda\pi)W^*(\rightarrow\tau\bar{\nu_\tau})$ may be written in terms of helicity amplitudes~\cite{Richman:1984gh,kutsckhe1996angular,Kadeer:2005aq} as
\begin{widetext}
\begin{align}
	\tilde I(\Omega_1,\Omega_2,\Omega_3)&=\sum_{m_1,m_1^\prime,\lambda_i,\lambda_j^\prime,s_3,s_3^\prime}\left\{ \rho_{ m_1 m_1^\prime}~D^{s_1}_{ m_1,\lambda_2-\lambda_3}(\Omega_1)D^{s_1*}_{ m_1^\prime,\lambda_2^\prime-\lambda_3^\prime}(\Omega_1)\nonumber\right. \left.H^*_{\lambda_2\lambda_3}H_{\lambda_2^\prime\lambda_3^\prime}~D^{s_2}_{\lambda_2,\lambda_4-\lambda_5}(\Omega_2)~D^{s_2*}_{\lambda_2^\prime,\lambda_4-\lambda_5}(\Omega_2)\right.\nonumber\\
    &\hspace{4cm} B^*_{\lambda_4\lambda_5}B_{\lambda_4\lambda_5}
	(-1)^{s_3+s_3^\prime}~\left.D^{s_3}_{\lambda_3,\lambda_6-\lambda_7}(\Omega_3)~D^{s_3^\prime*}_{\lambda_3^\prime,\lambda_6-\lambda_7}(\Omega_3)L^*_{\lambda_6\lambda_7}L_{\lambda_6\lambda_7}\right\}\,,\label{angdistD}
\end{align}
\end{widetext}
Here $\rho_{ m_1 m_1^\prime}$ is the density matrix for $\Lambda_b$ spin quantized along $\mathbf{z_1}$. 
We have
\begin{align}
	D^j_{m,m^\prime}(\Omega_i) &\equiv e^{-i m \phi_i}d^j_{m,m^\prime}(\theta_i)e^{i m^\prime \phi_i}~,
\end{align}
where $D^j_{m,m^\prime}(\Omega_i)$ is the Wigner's D-matrix, and  we have taken the Euler angles corresponding to $\Omega_i$ to be $(\phi_i,\theta_i,-\phi_i)$, using the Jacob-Wick convention~\cite{Jacob:1959at}. Note that $s_3$ can take the values $0$ and $1$. Further, $H_{\lambda_i\lambda_j}$ and $B_{\lambda_4\lambda_5}$ are the hadronic matrix elements in the helicity basis for $\Lambda_b\rightarrow \Lambda_c(\lambda_i) W^*(\lambda_j)$ and $\Lambda_c\rightarrow \Lambda(\lambda_4) \pi(\lambda_5)$ decays, respectively, and $L_{\lambda_6 \lambda_7}$ is the leptonic matrix element for $W^*\rightarrow \tau(\lambda_6) \nu_\tau(\lambda_7)$. 

We assume that the initial $\Lambda_b$ is unpolarized, and  sum over all possible spin components of $\Lambda_b$ as well as integrate over the angles $\theta_1$ and  $\phi_1$. This reduces the angular distribution to the form
\begin{widetext}
\begin{align}
	I(\Omega_2,\Omega_3)&=\sum_{\lambda_i,\lambda_j^\prime,s_3,s_3^\prime}\left\{2\pi\delta_{(\lambda_2-\lambda_3),(\lambda_2^\prime-\lambda_3^\prime)}(-1)^{s_3+s_3^\prime}~H^*_{\lambda_2\lambda_3}H_{\lambda_2^\prime\lambda_3^\prime}D^{s_2}_{\lambda_2,\lambda_4-\lambda_5}(\Omega_2)~D^{s_2*}_{\lambda_2^\prime,\lambda_4-\lambda_5}(\Omega_2)\nonumber\right.\\
	 &\left. B^*_{\lambda_4\lambda_5}B_{\lambda_4\lambda_5}\,D^{s_3}_{\lambda_3,\lambda_6-\lambda_7}(\Omega_3)~D^{s_3^\prime*}_{\lambda_3^\prime,\lambda_6-\lambda_7}(\Omega_3)L^*_{\lambda_6\lambda_7}L_{\lambda_6\lambda_7}\right\}\,.
\end{align}
\end{widetext}
Note that since the direction of axes $\mathbf{z_2}$ and $\mathbf{z_3}$ are opposite and the direction $\mathbf{x_2}=-\mathbf{x_3}$ is arbitrary, the individual values of $\phi_2$ and $\phi_3$ are arbitrary but the combination $\chi\equiv\phi_2+\phi_3$ is physical. This is indeed the physical angle between the decay planes of  $\Lambda_c\rightarrow \Lambda+\pi$ and $W^*\rightarrow \tau+\nu_\tau$. As a result, we get the angular distribution of the decay in terms of three physical angles: $\theta_2$, $\theta_3$ and $\chi$. From now on, we refer to $\theta_2$ as $\theta_c$ and $\theta_3$ as $\theta_l$. We assume that the decay $\Lambda_b\rightarrow\Lambda_c(\rightarrow\Lambda\pi)\tau\bar\nu_\tau$ is completely identified, so that the values of $\theta_c$, $\theta_l$ and $\chi$ are available for every event. Note that the three-momentum of $\tau$ can be reconstructed in the three-prong topology using the direction of $\tau$ track~\cite{LHCb:2022piu, LHCb:2017rln}.

\subsection{Hadronic and leptonic matrix elements}

The hadronic matrix elements for the $\Lambda_b\rightarrow\Lambda_c W^*$ transition, in the rest frame of $\Lambda_b$, are given by~\cite{Datta:2017aue}
\begin{align}
	H^V_{\lambda_2,\lambda_3}&=(1+g_L+g_R)\epsilon^{*\mu}(\lambda_3)\bra{\Lambda_c(\lambda_2)}\bar{c}\gamma_\mu b\ket{\Lambda_b}~,\nonumber\\
	H^A_{\lambda_2,\lambda_3}&=(1+g_L-g_R)\epsilon^{*\mu}(\lambda_3)\bra{\Lambda_c(\lambda_2)}\bar{c}\gamma_\mu \gamma_5 b \ket{\Lambda_b}~,\nonumber\\
	H^S_{\lambda_2,\lambda_3}&=g_S\bra{\Lambda_c(\lambda_2)}\bar{c} b\ket{\Lambda_b}~,\nonumber\\
	H^P_{\lambda_2,\lambda_3}&=g_P\bra{\Lambda_c(\lambda_2)}\bar{c}\gamma_5 b\ket{\Lambda_b}~.\label{hadronicdef}
\end{align}
Here $\epsilon^\mu$ represents the polarization vector of $W^*$. Only certain combination of these hadronic matrix elements have non-zero contributions to the decay amplitude. For the vector and axial-vector currents, non-zero contributions are seen arise only from the specific combination $H^{VA}_{\lambda_2,\lambda_3}=H^V_{\lambda_2,\lambda_3}-H^A_{\lambda_2,\lambda_3}$. Similarly, for the scalar and pseudoscalar currents, non-zero contributions come only from $H^{SP}_{\lambda_2,\lambda_3}=H^S_{\lambda_2,\lambda_3}+H^P_{\lambda_2,\lambda_3}$.
The explicit expressions of the hadronic amplitudes for different values of helicities~\cite{Datta:2017aue} are provided in Appendix \ref{app: hadronic} .

The amplitudes for the $\Lambda_c\rightarrow \Lambda\pi$ transition in the $\Lambda_c$ rest frame are defined as
\begin{equation}
	B_{\lambda_4\lambda_5}=\bra{\Lambda(\lambda_4)\pi(\lambda_5)}\ket{\Lambda_c(\lambda_2)}~.
\end{equation}
There are two independent amplitudes $B_{\frac{1}{2},0}$ and $B_{-\frac{1}{2},0}$. We do not use the analytic expressions for these amplitudes but calculate them in terms of the average decay rate $\Gamma_{\Lambda\pi}$ and the polarization asymmetry $\alpha_P$, defined as~\cite{Becirevic:2022bev}
\begin{align}
	\Gamma_{\Lambda\pi}&=\frac{\sqrt{Q}}{32\pi m_{\Lambda_c}^3}\left(|B_{\frac{1}{2},0}|^2+|B_{-\frac{1}{2},0}|^2\right),\\
	\alpha_P &=\frac{|B_{\frac{1}{2},0}|^2-|B_{-\frac{1}{2},0}|^2}{|B_{\frac{1}{2},0}|^2+|B_{-\frac{1}{2},0}|^2}~,
\end{align}
where $Q=(m_{\Lambda_c}^2-(m_\Lambda+m_\pi)^2)(m_{\Lambda_c}^2-(m_\Lambda-m_\pi)^2)$. The values of $\Gamma_{\Lambda\pi}$ and $\alpha_P$ are taken from measurements~\cite{ParticleDataGroup:2022pth}. 

The leptonic amplitudes for $W^*\rightarrow \tau\bar\nu_\tau$ are given as
\begin{align}
	L^{VA}_{\lambda_6,\lambda_7}&=\epsilon^\mu(\lambda_3)\bra{\tau\bar{\nu}_\tau}\bar{\tau}\gamma_\mu P_L \nu_{\tau}\ket{0}~,\\ L^{SP}_{\lambda_6,\lambda_7}&=\bra{\tau\bar{\nu}_\tau}\bar{\tau} P_L \nu_{\tau}\ket{0}~.
\end{align}
In the rest frame of $W^*$, these leptonic amplitudes are~\cite{Chang:2018sud}
\begin{align}
	L^{VA}_{-\frac{1}{2},\frac{1}{2}}&=\sqrt{8(q^2-m_{\tau}^2)}~,\\
	L^{VA}_{\frac{1}{2},\frac{1}{2}}&=\frac{ m_\tau}{\sqrt{2q^2}}\sqrt{8(q^2-m_{\tau}^2)}~,\\
	L^{SP}_{-\frac{1}{2},\frac{1}{2}}&=\sqrt{4(q^2-m_{\tau}^2)}~.
\end{align}
Here $q^2$ is the invariant mass of the $(\tau, \bar\nu_\tau)$ pair.

\subsection{The angular distribution} 

After summing over the helicities of all particles and integrating over the non-measurable angles, we obtain the normalized angular distribution in the form
\begin{align}
	&\frac{1}{(d\Gamma/dq^2)}\frac{d\Gamma}{dq^2\,d\cos\theta_c\,d\cos\theta_l\,d\chi}\nonumber\\
	&= A_0+A_1\,\cos \theta_c + A_2\,\cos\theta_l\nonumber\\
    &\quad+A_3\,\cos\theta_c\,\cos\theta_l +A_4\,\cos^2\theta_l\nonumber\\
    &\quad+A_5\,\cos\theta_c\cos^2\theta_l\nonumber\\
	&\quad+A_6\,\sin\theta_c\sin\theta_l\cos\chi\nonumber\\
    &\quad+A_7\,\sin\theta_c\sin\theta_l\sin\chi\nonumber\\
	&\quad+A_{8}\,\sin\theta_c\sin\theta_l\cos\theta_l\cos\chi\nonumber\\
	&\quad + A_{9}\,\sin\theta_c\sin\theta_l\cos\theta_l\sin\chi~.\label{angdist}
\end{align}
 The coefficients $A_0$ to $A_9$ in eq.\,(\ref{angdist}) can be extracted from a fit to the observed angular distribution or by the method of angular moments~\cite{Dighe:1998vk}. We have provided the weighting functions required to extract these coefficients in Appendix~\ref{app:moments}. 
The coefficients $A_0$ to $A_{9}$, which we refer to as the angular observables, are given in eqs.\,(\ref{A0})--(\ref{A9}), in terms of the total decay width
\begin{align}
\Gamma_0&\equiv\frac{2}{3}\left\{(m_\tau^2+2q^2)\left(|H^{VA}_{-\frac{1}{2},-1}|^2+|H^{VA}_{-\frac{1}{2},0}|^2+|H^{VA}_{\frac{1}{2},0}|^2\right.\right.\nonumber\\
&\left.\left.+|H^{VA}_{\frac{1}{2},1}|^2\right)+3\left(|m_\tau H^{VA}_{-\frac{1}{2},t}+\sqrt{q^2}\,H^{SP}_{-\frac{1}{2},t}|^2\right.\right.\nonumber\\
&\left.\left.+|m_\tau H^{VA}_{\frac{1}{2},t}+\sqrt{q^2}\,H^{SP}_{\frac{1}{2},t}|^2\right)\right\}~.
\end{align}

\begin{widetext}
\begin{align}
A_0&=\frac{1}{\Gamma_0}\left( \frac{1}{2}\left(m_{\tau}^2+q^2\right) | H^{VA}_{\frac{1}{2},1}| {}^2+\frac{1}{2} \left(m_{\tau}^2+q^2\right) | H^{VA}_{-\frac{1}{2},-1}| {}^2+| m_{\tau}H^{VA}_{\frac{1}{2},t} + \sqrt{q^2}\,H^{SP}_{\frac{1}{2},t}| {}^2\right.\nonumber\\
&\left.+|m_{\tau} H^{VA}_{-\frac{1}{2},t} + \sqrt{q^2}\,H^{SP}_{-\frac{1}{2},t}| {}^2+q^2 | H^{VA}_{\frac{1}{2},0}| {}^2+q^2 | H^{VA}_{-\frac{1}{2},0}| {}^2\right)~,\label{A0}\\
A_1&=\frac{\alpha_P}{2\Gamma_0}\left(\left(m_{\tau}^2+q^2\right) | H^{VA}_{\frac{1}{2},1}| {}^2+\left(m_{\tau}^2+q^2\right) | H^{VA}_{-\frac{1}{2},-1}| {}^2-2 |m_{\tau} H^{VA}_{\frac{1}{2},t} + \sqrt{q^2}\,H^{SP}_{\frac{1}{2},t}| {}^2\right.\nonumber\\
&\left.+2 | m_{\tau}H^{VA}_{-\frac{1}{2},t} + \sqrt{q^2}\,H^{SP}_{-\frac{1}{2},t}| {}^2-2 q^2 | H^{VA}_{\frac{1}{2},0}| {}^2+2 q^2 | H^{VA}_{-\frac{1}{2},0}| {}^2\right)~,\label{A1}\\
A_2&=-\frac{1}{\Gamma_0}\left(q^2 | H^{VA}_{\frac{1}{2},1}| {}^2+q^2 | H^{VA}_{-\frac{1}{2},-1}| {}^2\right.\nonumber\\
&\left.- 2\Re{ \left(m_{\tau}H^{VA}_{\frac{1}{2},0} (m_{\tau}H^{VA}_{\frac{1}{2},t}+\sqrt{q^2}\,H^{SP}_{\frac{1}{2},t}){}^*+m_{\tau}H^{VA}_{-\frac{1}{2},0} (m_{\tau}H^{VA}_{-\frac{1}{2},t}+\sqrt{q^2}\,H^{SP}_{-\frac{1}{2},t}){}^*\right)}\right)~,\\
A_3&=-\frac{\alpha_P}{\Gamma_0}\left(q^2 | H^{VA}_{\frac{1}{2},1}| {}^2+q^2 | H^{VA}_{-\frac{1}{2},-1}| {}^2\right.\label{A3}\nonumber\\
&\left.+2m_{\tau}\Re{ \left(H^{VA}_{\frac{1}{2},0} (m_{\tau}H^{VA}_{\frac{1}{2},t}+\sqrt{q^2}\,H^{SP}_{\frac{1}{2},t}){}^*-H^{VA}_{-\frac{1}{2},0} (m_{\tau}H^{VA}_{-\frac{1}{2},t}+\sqrt{q^2}\,H^{SP}_{-\frac{1}{2},t}){}^*\right)}\right)~,\\
A_4&=-\frac{1}{2\Gamma_0} \left(m_{\tau}^2-q^2\right) \left(| H^{VA}_{\frac{1}{2},1}| {}^2+| H^{VA}_{-\frac{1}{2},-1}| {}^2-2 \left(| H^{VA}_{\frac{1}{2},0}| {}^2+| H^{VA}_{-\frac{1}{2},0}| {}^2\right)\right)~,\\
A_5&=\frac{\alpha_P}{\Gamma_0}(m_{\tau}^2-q^2)\left(2| H^{VA}_{\frac{1}{2},0}| {}^2-| H^{VA}_{\frac{1}{2},1}| {}^2+| H^{VA}_{-\frac{1}{2},-1}| {}^2-2 | H^{VA}_{-\frac{1}{2},0}| {}^2\right)\label{A5}~,\\
A_6&=-\frac{\alpha_P}{\sqrt{2}\Gamma_0}2\textrm{Re}\left\{\left(H^{VA}_{-\frac{1}{2},-1}\right){}^* \left(m_{\tau}(m_{\tau}H^{VA}_{\frac{1}{2},t}+\sqrt{q^2}\,H^{SP}_{\frac{1}{2},t})+q^2 H^{VA}_{\frac{1}{2},0}\right)\right.\nonumber\\
&\left. ~~+\left(H^{VA}_{\frac{1}{2},1}\right){}^* \left(q^2 H^{VA}_{-\frac{1}{2},0}- m_{\tau}(m_{\tau}H^{VA}_{-\frac{1}{2},t}+\sqrt{q^2}\,H^{SP}_{-\frac{1}{2},t})\right)\right\}~,\\
A_7&=-\frac{\alpha_P}{\sqrt{2}\Gamma_0}2\textrm{Im}\left\{\left(H^{VA}_{\frac{1}{2},1}\right){}^* \left(q^2 H^{VA}_{-\frac{1}{2},0}- m_{\tau}(m_{\tau}H^{VA}_{-\frac{1}{2},t}+\sqrt{q^2}\,H^{SP}_{-\frac{1}{2},t})\right)\right.\nonumber\\
&\left.~~-\left(H^{VA}_{-\frac{1}{2},-1}\right){}^* \left( m_{\tau}(m_{\tau}H^{VA}_{\frac{1}{2},t}+\sqrt{q^2}\,H^{SP}_{\frac{1}{2},t})+q^2 H^{VA}_{\frac{1}{2},0}\right)\right\}~,\\
A_8&=\frac{\alpha_P}{\sqrt{2}\Gamma_0}(m_{\tau}^2-q^2)2\Re{\left(H^{VA}_{-\frac{1}{2},-1} \left(H^{VA}_{\frac{1}{2},0}\right){}^*-H^{VA}_{-\frac{1}{2},0} \left(H^{VA}_{\frac{1}{2},1}\right){}^*\right)}~,\\
A_9&=\frac{\alpha_P}{\sqrt{2}\Gamma_0}(m_{\tau}^2-q^2)2\Im \left(H^{VA}_{-\frac{1}{2},-1} \left(H^{VA}_{\frac{1}{2},0}\right){}^*+H^{VA}_{\frac{1}{2},1} \left(H^{VA}_{-\frac{1}{2},0}\right){}^*\right)\label{A9}~.
\end{align}
\end{widetext}
Note that while $A_7$ and $A_9$ change sign under  $g_R\leftrightarrow g_R^*$ (see the explicit expressions for $H^{VA}_{\lambda_2, \lambda_3}$ in Appendix \ref{app: hadronic}),  all the other angular observables are invariant under this interchange.

The mapping of the angular observables above to the ones calculated in earlier literature~\cite{Boer:2019zmp} has been provided in Appendix~\ref{app:mapping}. In the next section, we will discuss the utility of these angular observables for identifying physics  beyond SM and physics beyond SMEFT. 

\begin{table*}[t]
	\begin{center}
		\begin{tabular}{|c|c|c|}
			\hline
			Observable & Experimental value & SM value \\
			\hline
			$R_D$ & $0.357 \pm 0.029$~\cite{BaBar:2012obs, BaBar:2013mob, Belle:2015qfa, Belle:2019gij, HFLAV:2022pwe, HFLAV:2023sum} & $0.298 \pm 0.004$~\cite{MILC:2015uhg, Na:2015kha, Aoki:2016frl, Bigi:2016mdz, HFLAV:2022pwe, HFLAV:2023sum}\\
			\hline
			$R_D^*$ & $0.284 \pm 0.012$~\cite{BaBar:2012obs, BaBar:2013mob, Belle:2015qfa,Belle:2016ure, LHCb:2015gmp,LHCb:2017smo, LHCb:2017rln, Belle:2016dyj, Belle:2017ilt, HFLAV:2022pwe, HFLAV:2023sum} & $0.254 \pm 0.005$~\cite{Bernlochner:2017jka, Jaiswal:2017rve, Fajfer:2012vx, HFLAV:2022pwe, HFLAV:2023sum}\\
			\hline
			$R_{j/\psi}$ & $0.71 \pm 0.17 \pm 0.18$~\cite{LHCb:2017vlu} & $0.258 \pm 0.004$~\cite{Cohen:2018dgz,Harrison:2020nrv}\\
			\hline
			$P_\tau^{D^*}$ & $-0.38 \pm 0.51 \pm 0.21$~\cite{Belle:2016dyj, Belle:2017ilt} & $-0.497 \pm 0.013$~\cite{Tanaka:2012nw}\\
			\hline
			$F_L^{D^*}$ & $0.60 \pm 0.08 \pm 0.035$~\cite{Belle:2019ewo} & $0.46 \pm 0.04$\cite{Alok:2016qyh}\\
			\hline
		\end{tabular}
	\end{center}
	\caption{\label{table: obsvalues}Current experimental values and SM predictions of the observables $R_D$, $R_D^*$, $R_{j/\psi}$, $P_\tau^{D^*}$ and $F_L^{D^*}$.}
\end{table*}

\section{Observables and numerical results}\label{sec:results}

Each of the coefficients $A_0$ to $A_9$ in eq.\,(\ref{angdist}) can be measured if the information about the angles $\theta_c$, $\theta_l$, $\chi$ and the leptonic invariant mass-square $q^2$ are known. In this section, we consider four NP scenarios $g_L$, $g_R$, $s_L$ and $s_R$, 
 where only the corresponding complex NP parameter $g_L$, $g_R$, $s_L$ and $s_R$, respectively, is nonzero. Here we have defined 
 \begin{align}
	s_L&=(g_S-g_P)/2~~~ \textrm{and}~~s_R=(g_S+g_P)/2~.
\end{align}
First, we identify the parameter values in all scenarios that give a reasonable fit to the above five observables, i.e a $\chi^2$ less than a pre-decided value or a $p$-value greater than a pre-decided minimum. Later, we check if the predictions of the angular observables in the $g_R$ scenario can be mimicked by any of the allowed set of parameters in the other scenarios.

\subsection{Constraining NP parameters based on meson decay observations}

In order to obtain the allowed parameter values in each scenario, we perform a $\chi^2$ fit for the corresponding NP parameter. The $\chi^2$ is defined as
\begin{align}
	\chi^2 &\equiv \sum_{i,j} (O^{\textrm{th}}_i-O^{\textrm{exp}}_i)\, \textrm{Cov}_{ij}^{-1}\,  (O^{\textrm{th}}_j-O^{\textrm{exp}}_j)~.
\end{align}
The experimentally measured central values of $O_i^{\textrm{exp}}$ and their uncertainties $\Delta O_i^{\textrm{exp}}$ are shown in Table~\ref{table: obsvalues}. The theoretical values of these observables are calculated based on \cite{Iguro:2022yzr}. The covariance matrix $\rm{Cov}_{ij}$ is defined as
\begin{align}
	\textrm{Cov}_{ij} &\equiv \Delta O_i^{\textrm{exp}} \rho_{ij}\, \Delta O_j^{\textrm{exp}} + \Delta O_i^{\textrm{th}} \delta_{ij}\, \Delta O_j^{\textrm{th}}~,
\end{align}
where $\rho_{ij}$ is the correlation between $i^{\rm{th}}$ and $j^{\rm{th}}$ observables. For a pair of independent measurements $i$ and $j$, we have $\rho_{ij}=\delta_{ij}$. For the measurements of $i \equiv R_{D}$ and $j \equiv R_{D^*}$, we have taken $\rho_{ij}=-0.37$~\cite{HFLAV:2023sum}. We further ignore the theoretical uncertainties as they are much smaller compared to the experimental uncertainties~\cite{Iguro:2022yzr}.

For any fixed value of the complex parameters $g_L$, $g_R$, $s_L$ and $s_R$, we calculate the $\chi^2$ using five observables as listed in Table~\ref{table: obsvalues}. The goodness of fit for each of the parameter values may be expressed in terms of $\chi^2$, $p$-values, confidence levels or number of sigmas. Since we are determining the goodness of fit for each NP parameter value independently, the number of degrees of freedom (d.o.f.) is the same as the number of observables, i.e. five. For 5 d.o.f., the correspondence among these four measures of goodness of fit is given in Table~\ref{tab: chi2}. Thus, stating that an NP parameter value is allowed at $2\sigma$ or $95.45\%$ C.L. is equivalent to stating that its $p$-value $p > 0.0455$, or $\chi^2 < 11.31$.

\begin{table}[h!]
    \centering
    \begin{tabular}{|c|c|c|c|}
        \hline
         Sigma & C.L. & $p$-value & $\chi^2$ \\
        \hline
        $1\sigma$ & 68.3\% & 0.317 & 5.89 \\
        $1.64\sigma$ & 90\% & 0.10 & 9.24 \\
        $2\sigma$ & 95.45\% & 0.0455 & 11.31 \\
        \hline
    \end{tabular}
    \caption{\label{tab: chi2} Correspondence among the four measures of the goodness of fit, with 5 degrees of freedom.}
\end{table}

\begin{figure}
	\centering
	\includegraphics[width=\columnwidth]{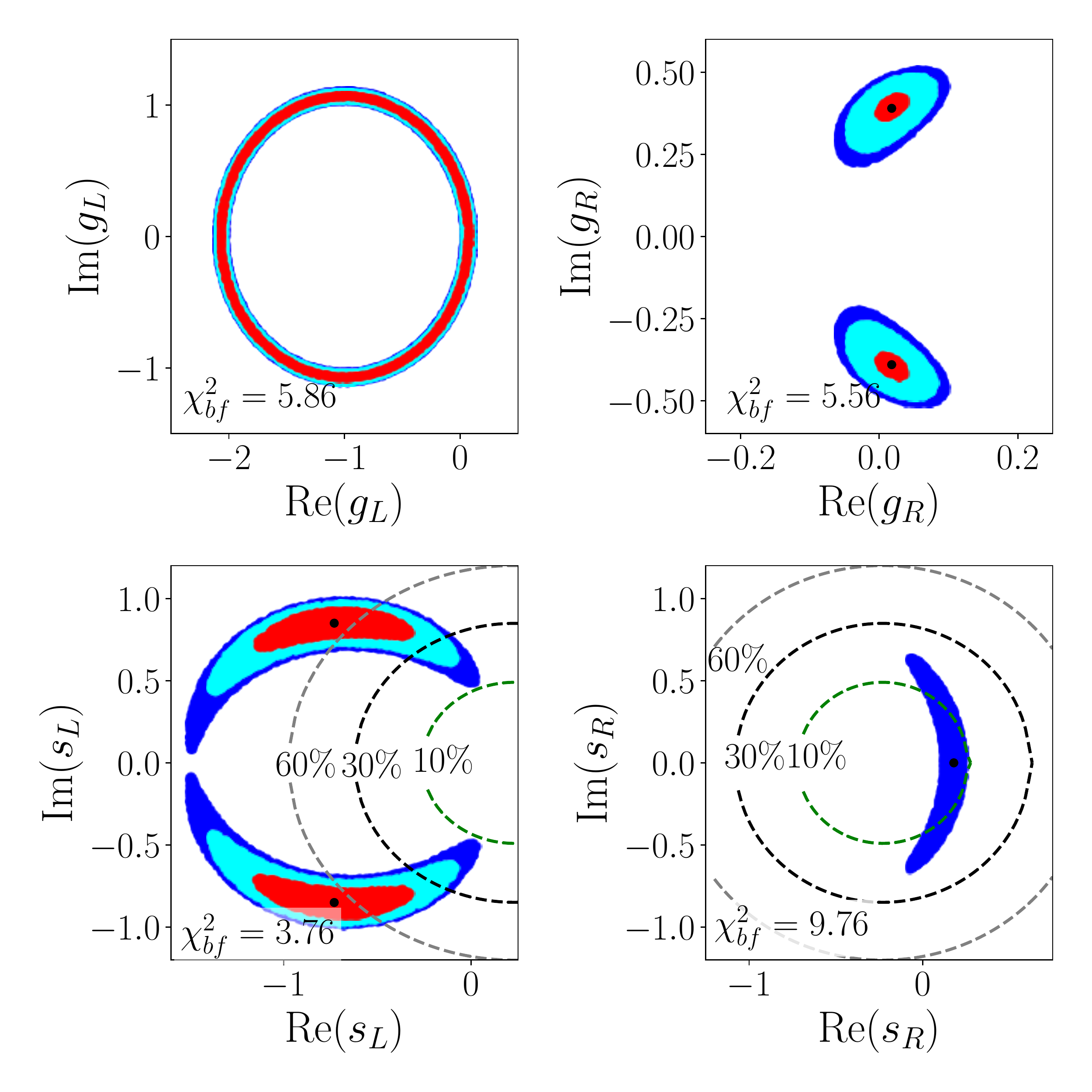}
	\caption{\justifying Regions allowed in the parameter space of $g_L$, $g_R$, $s_L$ and $s_R$ in the absence of any restriction on  $\mathcal{B}(B_c\rightarrow \tau \bar\nu_\tau)$. The red, cyan and blue regions are allowed at $1\sigma$, $1.64\sigma$ and $2\sigma$, respectively (5 d.o.f). The black dots represent the best-fit values of the NP parameters.  Note that the $\chi^2$ values in the $g_L$ scenario are degenerate along the red circular region, and there is no single best-fit value. The dashed (gray, black, green) contours indicate the allowed values of $s_L$ and $s_R$ corresponding to the upper bound $\mathcal{B}(B_c\rightarrow \tau \bar\nu_\tau)< (60\%,\,30\%,\,10\%)$. For $g_L$ and $g_R$, even the contours for $\mathcal{B}(B_c\rightarrow \tau \bar\nu_\tau)< 10\%$ are outside the range shown in the figure. }
	\label{fig: regions} 
\end{figure}

In Fig.~\ref{fig: regions}, we show the NP parameter values allowed at $1\sigma$, $1.64\sigma$ and $2\sigma$ in the four scenarios under consideration. The best-fit points in each scenario, calculated using the \textit{python} package `iminuit'~\cite{iminuit}, have been indicated in this figure as well as in Table~\ref{tab: bf2}. It is observed that in the scenarios $g_L$, $g_R$ and $s_L$, there are parameter values allowed at $1\sigma$. However, no parameter values in the $s_R$ scenario are allowed even at $1.64\sigma$, though there are some values allowed at $2\sigma$.
The SM itself has a $p$-value of 0.00045 and is disfavored at more than 3.5$\sigma$.

\begin{table*}
    \centering
    \renewcommand{\arraystretch}{1.1}
    \begin{tabular}{|c|c|c|c||c|c|c||c|c|c|}
        \hline
         \multirow{2}{*}{Scenario} & \multicolumn{3}{c||}{$\mathcal{B}(B_c\to \tau \bar\nu_\tau)$ unrestricted}& \multicolumn{3}{c||}{$\mathcal{B}(B_c\to \tau \bar\nu_\tau) < 30\%$} & \multicolumn{3}{c|}{$\mathcal{B}(B_c\to \tau \bar\nu_\tau) < 10\%$}\\
         \cline{2-10}
         & best-fit & $\chi^2_{bf}$ & $p$-value$_{(bf)}$ &best-fit & $\chi^2_{bf}$ & $p$-value$_{(bf)}$ &best-fit & $\chi^2_{bf}$ & $p$-value$_{(bf)}$\\
         \hline
         SM & - & 22.35 & 0.00045 & - & 22.35 & 0.00045 & - & 22.35 & 0.00045\\
         \hline
         $g_L$ & \textasteriskcentered & 5.86 &0.32 & \textasteriskcentered & 5.86 &0.32 & \textasteriskcentered & 5.86 &0.32\\
         \hline
         $g_R$ & $0.018 \pm 0.39i$ & 5.56 & 0.35 & $0.018 \pm 0.39i$ & 5.56 & 0.35 & $0.018 \pm 0.39i$ & 5.56 & 0.35 \\
         \hline 
         $s_L$ & $-0.73 \pm 0.85i$ & 3.76 & 0.58 & $-0.22-0.72i$ & 7.24 & 0.20 & $0.04 -0.45i$ & 11.78 & 0.04\\
         \hline
         $s_R$ & $0.18+0.00i$ & 9.76 & 0.08 & $0.18+0.00i$ & 9.76 & 0.08 & $0.18+0.00i$ & 9.76 & 0.08 \\
         \hline
    \end{tabular}
    \caption{\justifying\label{tab: bf2} Best-fit values of the NP parameters for the scenarios $g_L$, $g_R$, $s_L$ and $s_R$, along with the corresponding $\chi^2$ and $p$-values (5 d.o.f) for these best-fit points, for unrestricted $\mathcal{B}(B_c\rightarrow \tau \bar\nu_\tau)$ (left), for $\mathcal{B}(B_c\rightarrow \tau \bar\nu_\tau)< 30\%$ (center) and for $\mathcal{B}(B_c\rightarrow \tau \bar\nu_\tau)< 10\%$ (right). Note that restricting ${\cal B}(B_c \to \tau \bar\nu_\tau)$ does not affect the best-fit points of $g_L$, $g_R$ and $s_R$. The star for the best-fit value in the $g_L$ scenario indicates that there is no single best-fit point (see Fig.~\ref{fig: regions}).}
\end{table*}

So far, we have not considered the observable $\mathcal{B}(B_c\rightarrow \tau \bar\nu_\tau)$, which is very sensitive to the scalar NP operators~\cite{Blanke:2019qrx}. The SM value for $\mathcal{B}(B_c\rightarrow \tau \bar\nu_\tau)$ is $\simeq 2.2\%$~\cite{ParticleDataGroup:2022pth}. Experimentally $B_c\rightarrow \tau \bar\nu_\tau$ branching fraction is not yet measured. In \cite{Alonso:2016oyd}, an upper bound $\mathcal{B}(B_c\rightarrow \tau \bar\nu_\tau)< 30\%$ is evaluated from the missing decay width of $B_c$ meson. On the other hand, an upper bound $\mathcal{B}(B_c\rightarrow \tau \bar\nu_\tau)< 10\%$ is obtained in~\cite{Akeroyd:2017mhr} using LEP data. However, it has been pointed out~\cite{Blanke:2018yud, Blanke:2019qrx, Bardhan:2019ljo} that the  $p_T$ dependence of the fragmentation function $b\rightarrow B_c$ has been overlooked in \cite{Akeroyd:2017mhr} and thus the bound may be overestimated. In \cite{Blanke:2018yud}, the authors suggest a more conservative bound $\mathcal{B}(B_c\rightarrow \tau \bar\nu_\tau)< 60\%$. We show the contours corresponding to the upper bounds $\mathcal{B}(B_c\rightarrow \tau \bar\nu_\tau)< 60\%\,,~30\%\,,~\rm{and}~10\%$ in figure \ref{fig: regions} to indicate the  values of $s_L$ and $s_R$ which would remain valid even with these constraints. These fit results match reasonably with earlier literature~\cite{Iguro:2022yzr, Das:2023gfz} and with the recent fit~\cite{Arslan:2023wgk} where $g_L$, $s_L$ and $s_R$ scenarios were analyzed. In future, the FCC-ee and CEPC  experiments expect to directly measure  $\mathcal{B}(B_c\rightarrow \tau \bar\nu_\tau)$ at the $\mathcal{O}(1\%)$ level~\cite{Zheng:2020ult, Amhis:2021cfy}. In our analysis, we first present our results with the condition $\mathcal{B}(B_c\rightarrow \tau \bar\nu_\tau)< 30\%$, and later we see the impact of a stronger bound $\mathcal{B}(B_c\rightarrow \tau \bar\nu_\tau)< 10\%$.

It is not possible in SMEFT to have as large a value of $g_R$ as mentioned in Table \ref{tab: bf2}. If such a large value of $g_R$ is measured experimentally, it would imply physics beyond SMEFT and hence a non-linear realization of the EW symmetry. In the following, we see how such a value of $g_R$ can be identified from the angular observables and how its effect can be distinguished from the other NP parameters.

\subsection{Angular observables in $\Lambda_b \to \Lambda_c(\to \Lambda \pi)\tau\bar\nu_\tau$ for BSM and beyond-SMEFT signatures}\label{obs}

Our aim is to look for confirmed signals of any kind of NP, and to further identify whether this NP needs terms beyond SMEFT for its description. We proceed in two steps. 

In the first step, we check for each of the observables whether the effects of $g_R$ at its best-fit value can be distinguished from the SM predictions.  The $2\sigma$ uncertainties on the observables due to the hadronic form factors are incorporated using the z-expansion coefficients and their correlated uncertainties as given in~\cite{Detmold:2015aaa}. We also include the effect of the uncertainty in the measurement of the polarization asymmetry $\alpha_P$~\cite{ParticleDataGroup:2022pth}.

In the second step, in order to distinguish the effects of $g_R$ from the effects of $g_L$, $s_L$ and $s_R$, we vary these NP parameters within their $2\sigma$-allowed ranges, while continuing to include the uncertainties due to the hadronic form factors and $\alpha_P$.  When the band of $g_R$ is non-overlapping with the bands of other scenarios in some $q^2$ range, there is potential for identification of the $g_R$ scenario.


\begin{figure}[t]
	\centering
	\includegraphics[width=\linewidth]{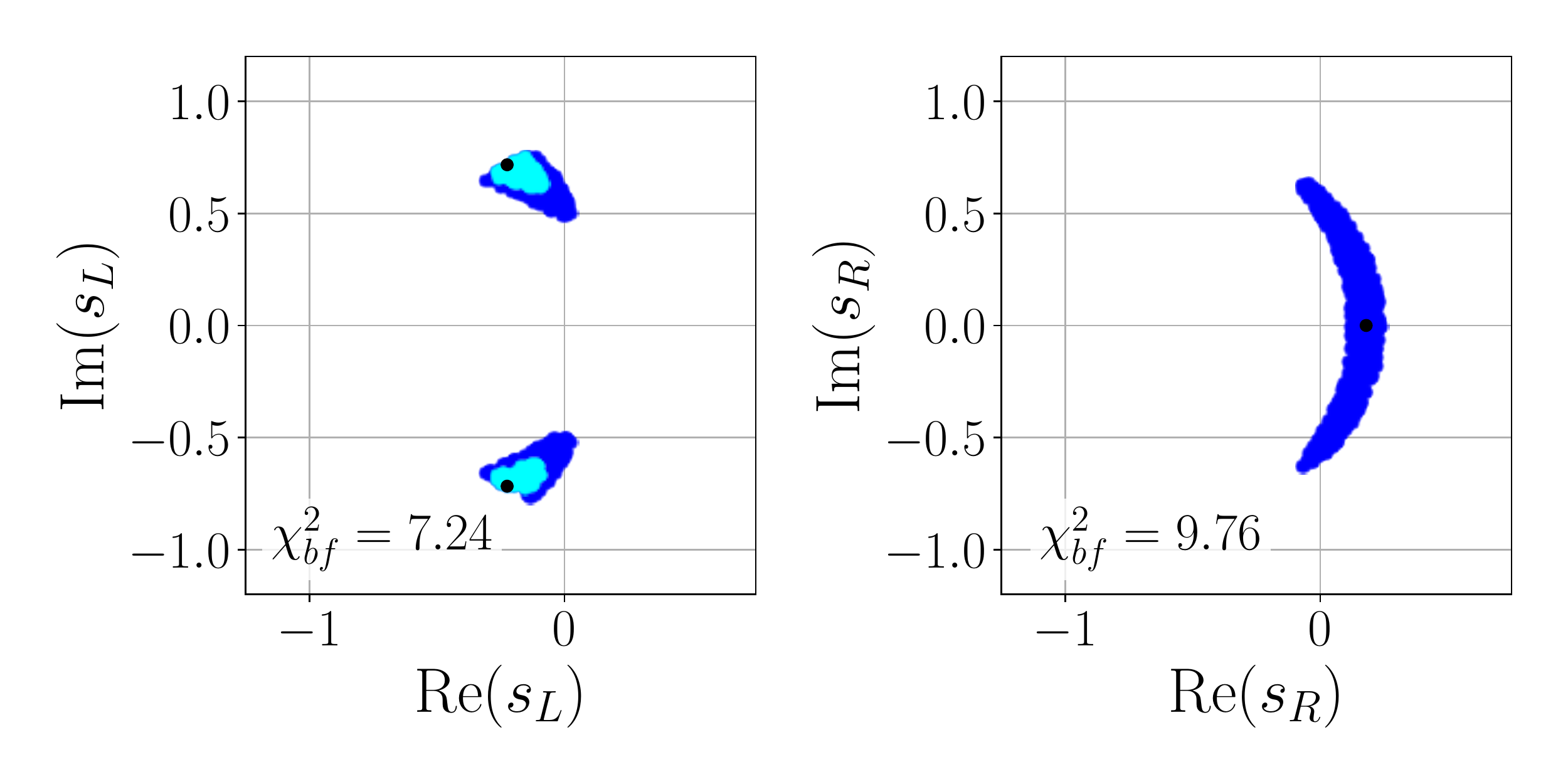}
	\caption{\justifying Allowed regions in the parameter spaces of  $s_L$ and $s_R$, with the upper bound $\mathcal{B}(B_c\rightarrow \tau \bar \nu_\tau) < 30\%$. The cyan and blue regions are allowed at $1.64\sigma$ and $2\sigma$, respectively (5 d.o.f).}
	\label{fig: regionsBr30} 
\end{figure}


In this subsection, we take the upper bound on  $\mathcal{B}(B_c\rightarrow \tau \bar \nu_\tau)$ to be a sharp cut off at $30\%$, i.e, we calculate the $\chi^2$ values for only those points in the parameter space for which $\mathcal{B}(B_c\rightarrow \tau \bar \nu_\tau) < 30\%$. It is observed that there is no change in the allowed regions for $g_L$, $g_R$ and $s_R$ compared to figure \ref{fig: regions}. However, for $s_L$, the allowed region gets severely restricted and no value of $s_L$ is allowed at $1\sigma$.  The allowed regions in the $s_L$ and $s_R$ parameter spaces are shown in figure \ref{fig: regionsBr30}. Note that with this upper bound, the $g_R$ scenario gives a better $\chi^2$ fit than the other scenarios. In particular, $\chi^2_{bf}(g_R)<\chi^2_{bf}(s_L)$, which was not the case for unrestricted $\mathcal{B}(B_c\rightarrow \tau \bar \nu_\tau)$.

We now proceed to calculate the decay width and the angular observables in $\Lambda_b \to \Lambda_c(\to \Lambda \pi)\tau\bar\nu_\tau$ as functions of $q^2$, and compare the predictions of the NP scenarios $g_L$, $g_R$, $s_L$ and $s_R$.

\begin{center}
    {\bf Decay width $d\Gamma/dq^2$}
\end{center}

\begin{figure}
	\centering
		\begin{subfigure}[b]{0.48\textwidth}
			\includegraphics[width=0.8\textwidth]{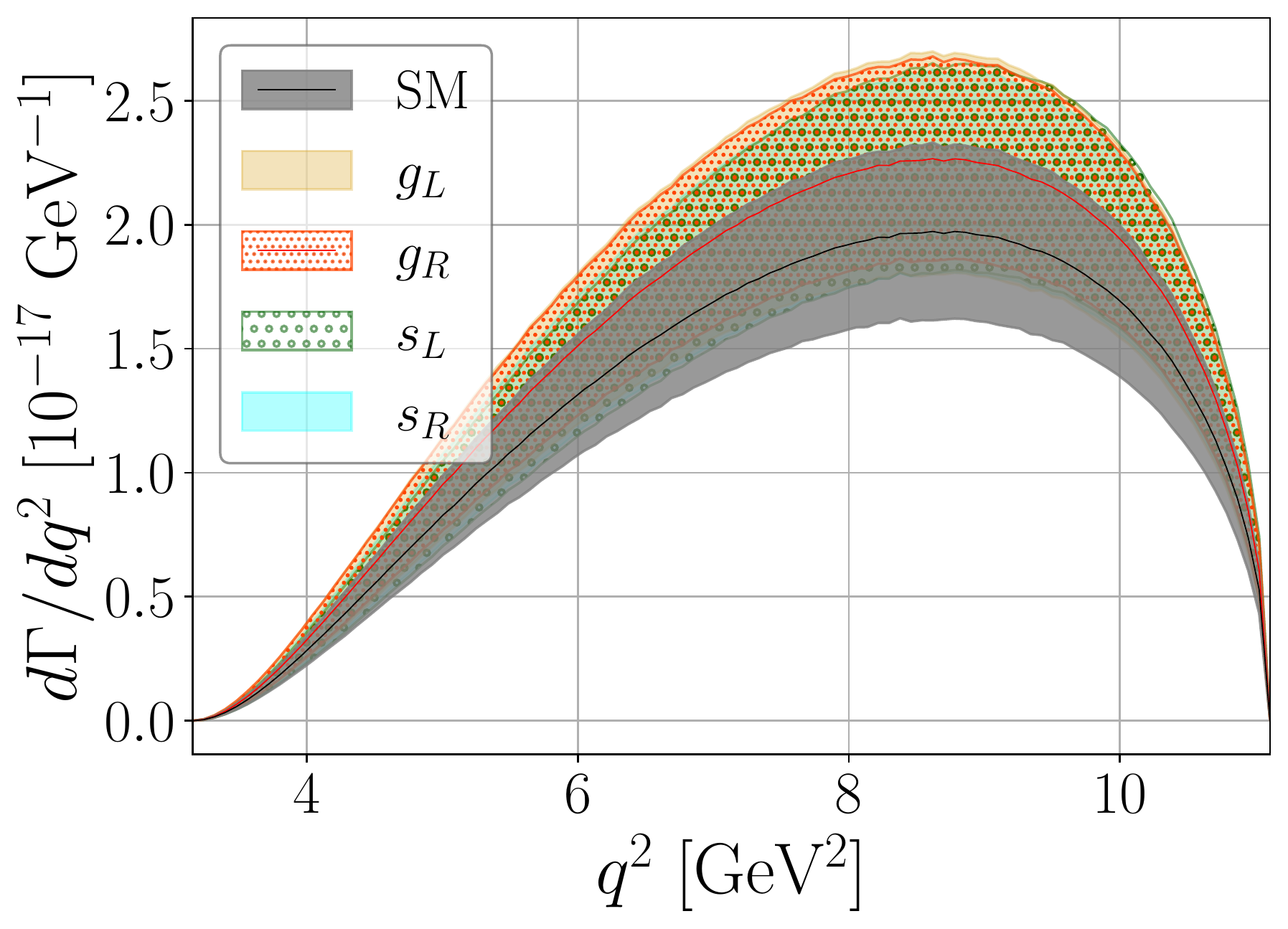}
		\end{subfigure}
	\hspace{0.2cm}
		\begin{subfigure}[b]{0.48\textwidth}
			\includegraphics[width=0.8\textwidth]{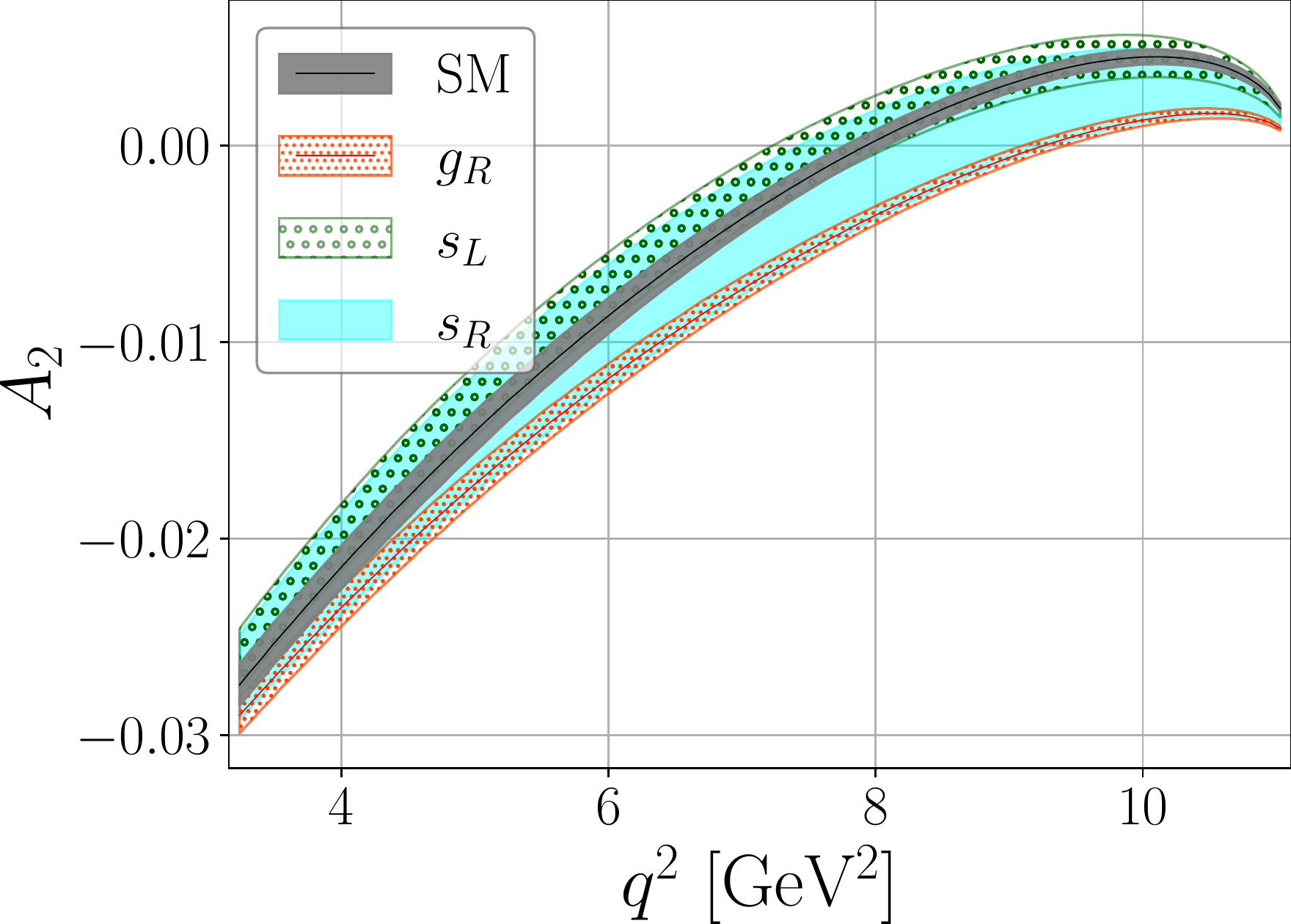}
		\end{subfigure}
		\caption{\justifying Top panel: The differential decay rate $d\Gamma/dq^2$ as a function of $q^2$, with SM and the NP scenarios $g_L$, $g_R$, $s_L$ and $s_R$. Bottom panel: The angular observable $A_2$ (the forward-backward asymmetry) as a function of $q^2$, with SM and the NP scenarios  $g_R$, $s_L$ and $s_R$. For both the observables, the values of $g_L$, $s_L$ and $s_R$ are varied within their $2\sigma$-allowed ranges, while $g_R$ is kept fixed at its best-fit value. For each scenario, $2\sigma$ uncertainties due to the hadronic form factors and the polarization asymmetry $\alpha_P$ have been included. The calculation of $d\Gamma/dq^2$ also includes the uncertainty in $V_{cb}$. The bound $\mathcal{B}(B_c\rightarrow \tau \bar \nu_\tau)<30\%$ is imposed for both these plots.}
		\label{A2fig}
	\end{figure}

In the top panel of figure~\ref{A2fig}, we show the effects of $g_L$, $g_R$, $s_L$ and $s_R$ on $d\Gamma/dq^2$. The best-fit value of $g_R$ and the $2\sigma$-allowed values of  $g_L$, $s_L$ and $s_R$ are taken. The $2\sigma$ uncertainties due to the hadronic form factors, $\alpha_P$ and $V_{cb}$ are included. We use the exclusive measurement of $V_{cb}$~\cite{ParticleDataGroup:2022pth}. Note that it will be almost impossible to distinguish $g_R$ from the SM or other NP scenarios, using only $d\Gamma/dq^2$.

\begin{center}
    {\bf Angular observables}
\end{center}

We divide the ten angular observables $A_0$,...,$A_9$  in five subsets: (i) $A_0$ and $A_4$, (ii) $A_2$, (iii) $A_1$, $A_3$ and $A_5$, (iv) $A_6$ and $A_8$, (v) $A_7$ and $A_9$. We then explore the effects of the NP parameters $g_L$, $g_R$, $s_L$ and $s_R$ on these observables, as functions of $q^2$. Note that when the NP appears only through $g_L$, eq.\,(\ref{HVA}) indicates that all the hadronic matrix elements, $H^{VA}_{\lambda,\lambda^\prime}$ get multiplied by a common factor $(1+g_L)$. As a result, in the angular observables given in eqs.\,(\ref{A0}--\ref{A9}), the dependence on $g_L$ cancels out completely. Consequently, the angular observables will not be affected by $g_L$. Therefore, in the following sections we only focus on the effects of $g_R$, $s_L$ and $s_R$. Note that the uncertainty in $V_{cb}$ gets canceled out in all the angular observables.
	
\paragraph*{\bf (i) $A_0$ and $A_4$: }
These two angular observables are the coefficients of the constant term and $\cos^2\theta_l$ in eq.\,(\ref{angdist}), respectively. They are trivially related as $A_0+A_4/3=1/2$. Even though they are easier to measure experimentally, we find that they cannot distinguish the effects of non-zero $g_R$ from SM when we include the uncertainties due to hadronic form factors. 

\paragraph*{\bf (ii) $A_2$: }

This is the coefficient of the term $\cos\theta_l$ in eq.\,(\ref{angdist}). This observable corresponds to the forward-backward asymmetry $A_{\rm{FB}}$ with respect to $\theta_l$. The bottom panel of figure~\ref{A2fig} shows the effects of $g_R$, $s_L$ and $s_R$ on $A_2$. As can be observed from the figure, the effects of $g_R$ can be distinguished from SM and $s_L$-scenario if $g_R$ takes its best-fit value. However, the $g_R$-scenario overlaps with the $s_R$-scenario, which means that the effect of the best-fit $g_R$ value cannot be distinguished from that of $s_R$. Note that the zero-crossing of $A_2$ shifts to higher values of $q^2$ compared to the SM in the $g_R$ scenario, as has been observed earlier for $A_{\rm{FB}}$~\cite{Mu:2019bin, Ray:2018hrx}.

\paragraph*{\bf (iii) $A_1$, $A_3$ and $A_5$: }

\begin{figure}
	\centering
	\begin{subfigure}[b]{0.48\textwidth}
		\includegraphics[width=0.8\textwidth]{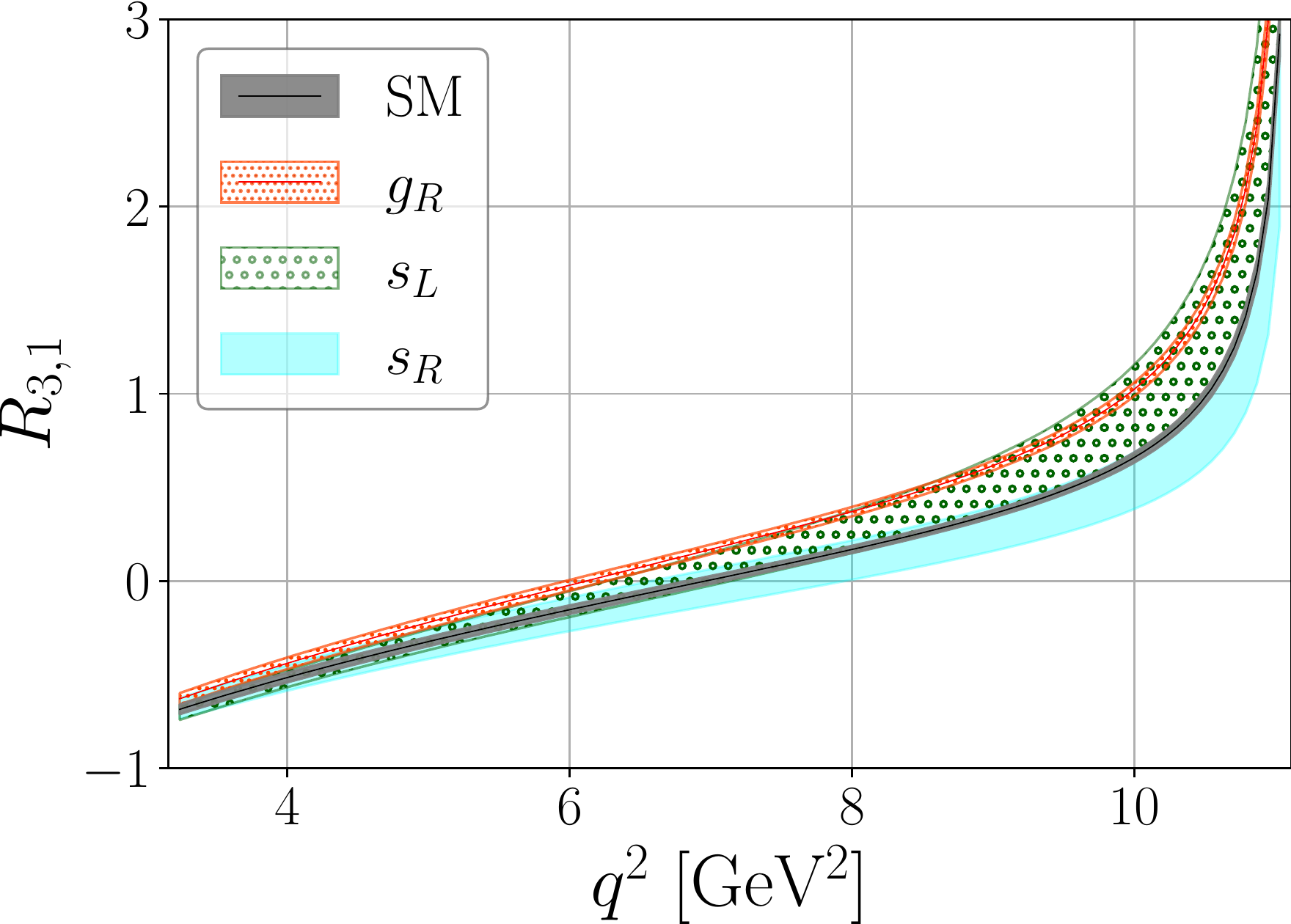}
	\end{subfigure}
	\hspace{0.3cm}
	\begin{subfigure}[b]{0.48\textwidth}
		\includegraphics[width=0.8\textwidth]{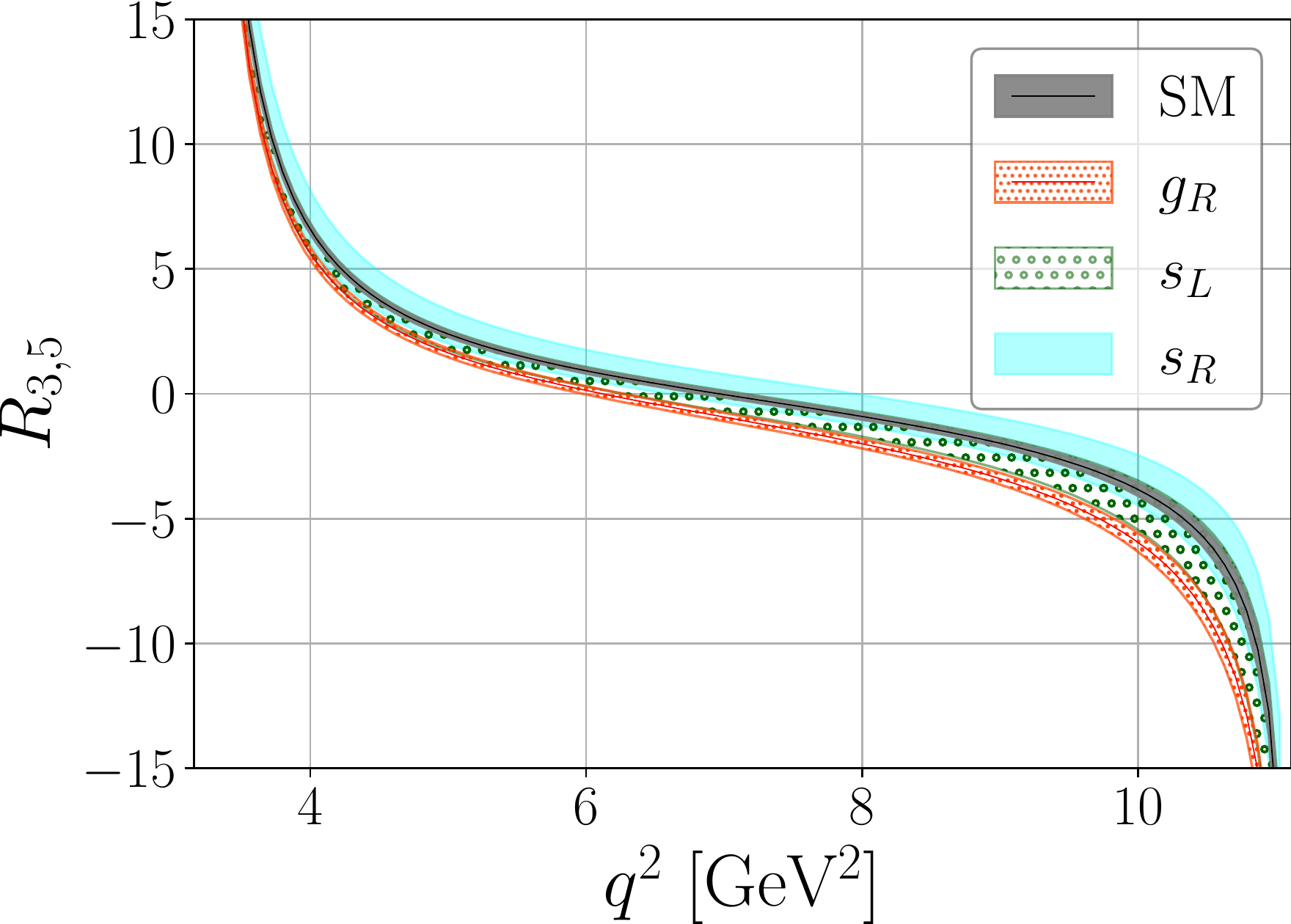}
	\end{subfigure}
	\caption{\justifying The ratios $R_{3,1}$ (top panel) and $R_{3,5}$ (bottom panel) as functions of $q^2$, with SM and the NP scenarios $g_R$, $s_L$ and $s_R$.  The values of $s_L$ and $s_R$ are varied within their $2\sigma$-allowed ranges, while $g_R$ is kept fixed at its best-fit value. For each scenario, $2\sigma$ uncertainties due to the hadronic form factors and the polarization asymmetry $\alpha_P$ have been included.  The bound $\mathcal{B}(B_c\rightarrow \tau \bar \nu_\tau)<30\%$ is imposed for both these plots.}
	\label{A135}
\end{figure}

These are the coefficients of the angular functions which are linear in $\cos\theta_c$. These angular observables are linear in the polarization asymmetry $\alpha_P$ as can be seen from eqs.\,(\ref{A1}), (\ref{A3}) and (\ref{A5}). Due to the large uncertainty in the measured value $\alpha_P=-0.84\pm 0.09$~\cite{ParticleDataGroup:2022pth},  $A_1$, $A_3$ and $A_5$ individually cannot distinguish the effects of $g_R$ from those of the other scenarios. Taking the ratios $R_{3,1}\equiv A_3/A_1$ and $R_{3,5}\equiv A_3/A_5$ cancels the uncertainty due to $\alpha_P$. These ratios would help in distinguishing the $g_R$ scenario from SM at higher $q^2$ values. This may be seen in figure~\ref{A135}. However, these ratios would not be able to distinguish $g_R$ from the other NP scenarios. 

\paragraph*{\bf (iv) $A_6$ and $A_8$: }

\begin{figure}
	\centering
	\begin{subfigure}[b]{0.48\textwidth}
		\includegraphics[width=0.8\textwidth]{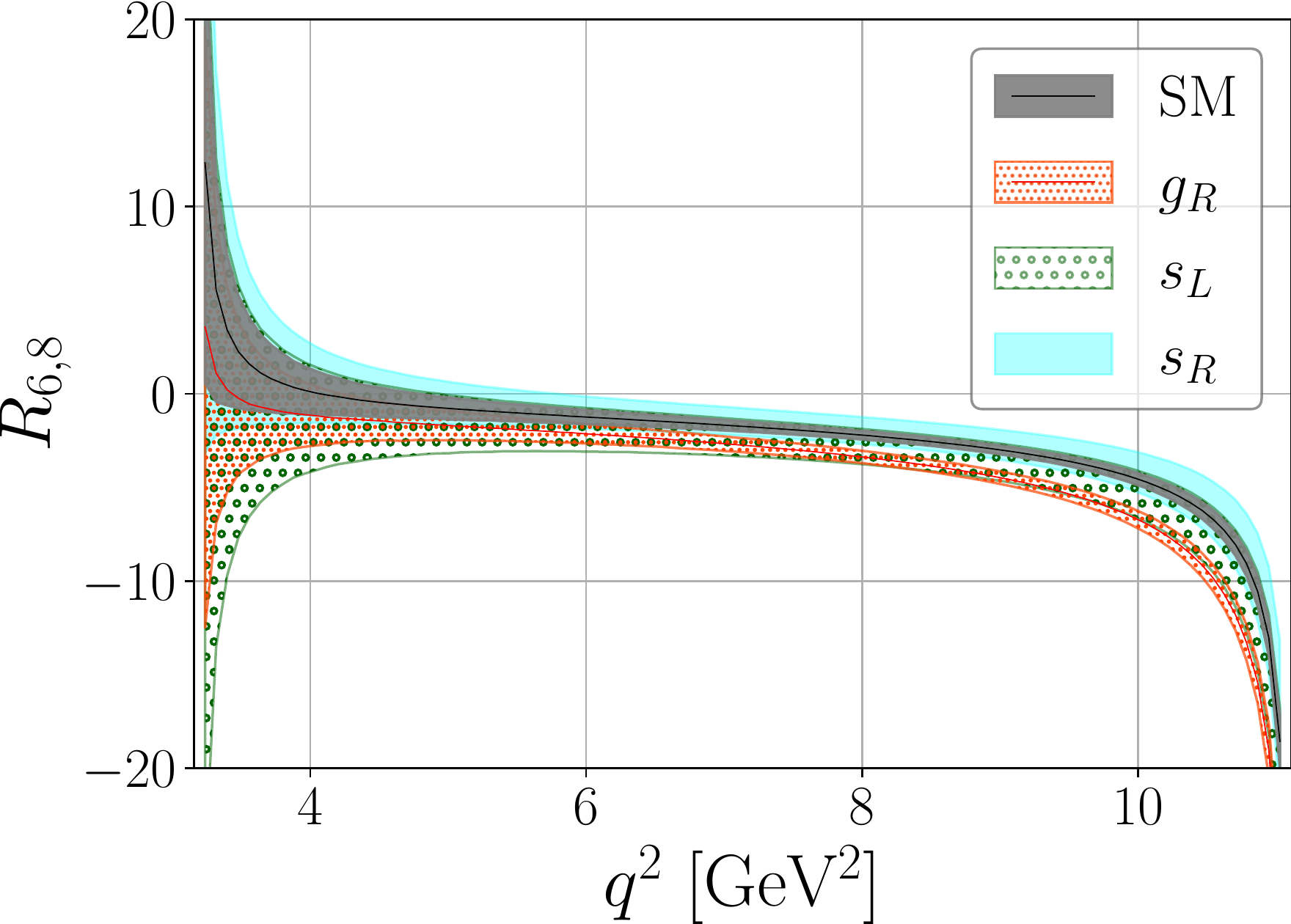}
	\end{subfigure}
	\hspace{0.2cm}
	\begin{subfigure}[b]{0.48\textwidth}
		\includegraphics[width=0.8\textwidth]{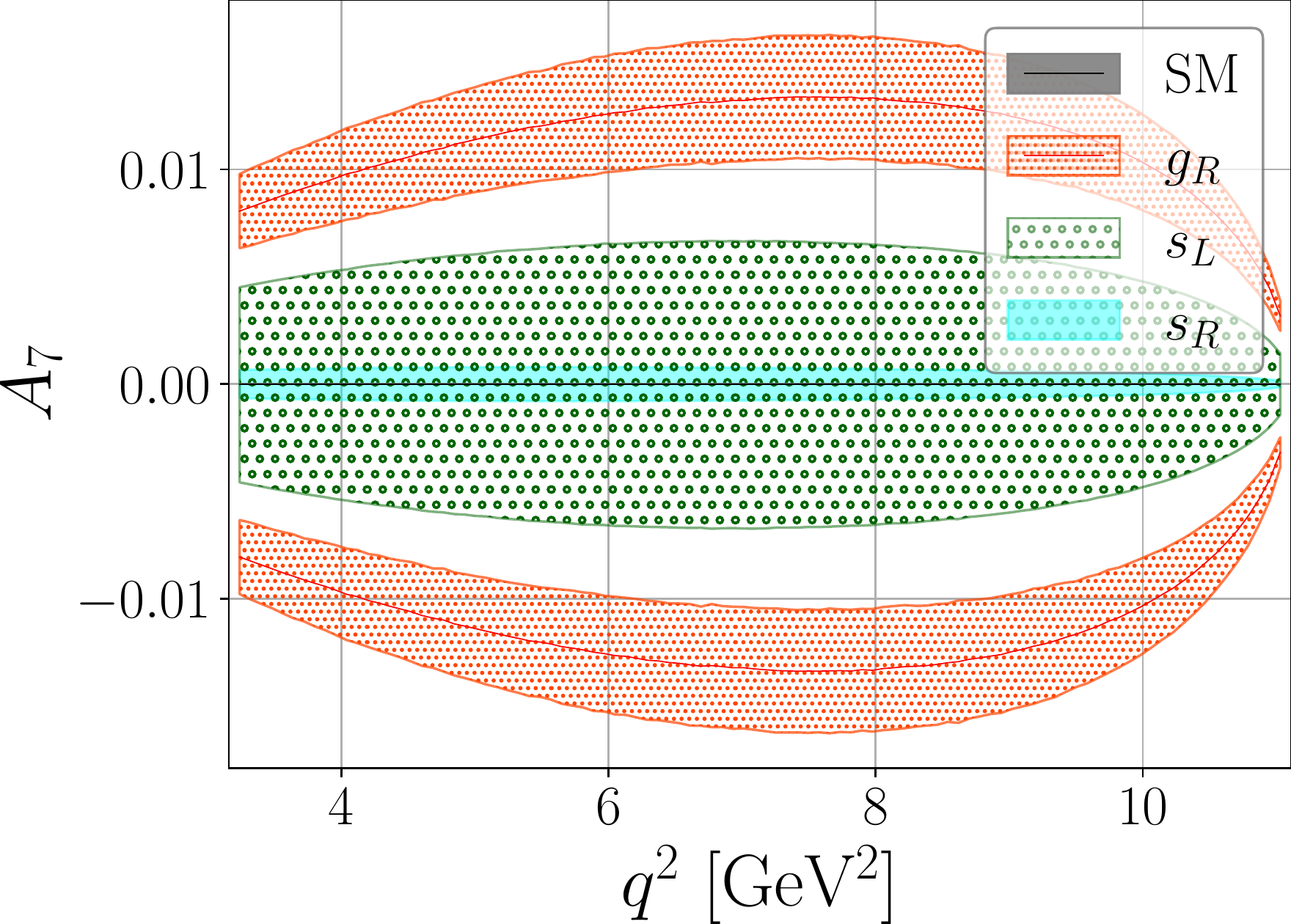}
	\end{subfigure}
	\caption{\justifying The ratios $R_{6,8}$ (top panel) and the angular observable $A_7$ (bottom panel) as functions of $q^2$, with SM and the NP scenarios $g_R$, $s_L$ and $s_R$.  The values of $s_L$ and $s_R$ are varied within  their $2\sigma$-allowed ranges, while $g_R$ kept fixed at its best-fit value as given in Table \ref{tab: bf2}. For each scenario, $2\sigma$ uncertainties due to the hadronic form factors and the polarization asymmetry $\alpha_P$ have been included.  The bound $\mathcal{B}(B_c\rightarrow \tau \bar \nu_\tau)<30\%$ is imposed for both these plots.}
	\label{A678}
\end{figure}

The angular observables $A_6$ and $A_8$ are the coefficients of the angular terms linear in $\cos\chi$. These observables are also linear in $\alpha_P$, and hence  individually they are not very suitable for identifying the effects of $g_R$. We construct the ratio $R_{6,8}\equiv A_6/A_8$ and show its dependence on NP parameters in the top panel of  figure~\ref{A678}. This ratio would only marginally help in separating the effects of $g_R$ from SM at higher $q^2$ values, and would not be able to distinguish from the other NP scenarios.


\paragraph*{\bf (v) $A_7$ and $A_9$:}

These observables are associated with angular functions which are linear in $\sin\chi$ and interchange sign under $g_R\leftrightarrow g_R^*$. We find that $A_9$ is identically zero in SM as well as in all NP scenarios considered here. However, $A_7$, which vanishes in SM, can be significantly non-zero in NP scenarios. In particular, the $g_R$ scenario, at its best-fit point as given in Table \ref{tab: bf2}, can give a non-zero value of $A_7$ that can be distinguished from the SM as well as from the $s_L$ and $s_R$ scenarios,  as shown in  figure~\ref{A678}. Note that $A_7$ is the only angular observable sensitive to the sign of Im$(g_R)$.

We summarize the discussion in this section in Table~\ref{chart}, where we show the effectiveness of the angular observables and their ratios in distinguishing the effects of the best-fit $g_R$ from SM and from the NP scenarios $g_L$, $s_L$ and $s_R$. If the band of predicted values of an observable for the best-fit value of $g_R$ does not overlap with the  band corresponding to any of the other scenarios in some $q^2$ range, we put a tick-mark in the corresponding cell. If the non-overlapping $q^2$ range is very small, then we add a cross in bracket.  If the $g_R$ prediction band always overlaps with the other scenario, we put only a cross in the corresponding cell. 

For $\mathcal{B}(B_c\rightarrow \tau \bar\nu_\tau)<30\%$, Table~\ref{chart} indicates:
\begin{itemize}
 \item The observable $A_2$ can distinguish the effects of the best-fit value of $g_R$ from the SM as well as from the NP scenarios $g_L$ and $s_L$, but not from $s_R$.
 \item The ratios $R_{3,1}$, $R_{3,5}$ and $R_{6,8}$ can distinguish the $g_R$ scenario from SM and $g_L$,  while failing to distinguish it from the $s_L$ or $s_R$ scenario. \item Only the angular observable $A_7$ can cleanly distinguish the $g_R$ scenario from the SM as well as from all the other NP scenarios.
\end{itemize}


\begin{table*}
	\begin{center}
			\begin{tabular}{| c || p{1.5cm} | p{1.0cm} | p{1.0cm} ||  p{1.5cm} | p{1cm} | p{1cm} |}
				\hline
				\multirow{2}{*}{\diagbox{Observable}{Scenario}}
				& \multicolumn{3}{c||}{$\mathcal{B}(B_c\rightarrow \tau \bar \nu_\tau)<30\%$}&\multicolumn{3}{|c|}{$\mathcal{B}(B_c\rightarrow \tau \bar \nu_\tau)<10\%$}\\
				\cline{2-7}  & SM , $g_L$ & $s_L$ & $s_R$ &  SM , $g_L$ & $s_L$ & $s_R$\\
				\hline
				$d\Gamma/dq^2$ & $\times$ & $\times$ & $\times$&$\times$ & $\checkmark$ & $\times$\\
				\hline
				$A_0$&$\times$ &$\times$ & $\times$&$\times$ &$\checkmark$ & $\times$\\
				\hline
				$A_1$&$\times$&$\times$ & $\times$ & 	$\times$&$\checkmark$ & $\times$\\ 
				\hline
				$A_2$&$\checkmark$ &$\checkmark^{(\times)}$& $\times$ &$\checkmark$ &$\checkmark$& $\times$ \\ 
				\hline
				$A_3$&$\checkmark^{(\times)}$ &$\times$ &$\times$&$\checkmark^{(\times)}$ &$\checkmark$& $\checkmark^{(\times)}$\\ 
				\hline
				$A_4$&$\times$ & $\times$ & $\times$& $\times$ & $\checkmark$ & $\times$\\ 
				\hline
				$A_5$&$\times$ &$\times$& $\times$ & $\times$ & $\checkmark$ & $\times$\\ 
				\hline
				$A_6$&$\times$ &$\times$& $\times$ & $\times$ & $\checkmark$ & $\times$\\ 
				\hline
				$A_7$&$\checkmark$ &$\checkmark$& $\checkmark$ & $\checkmark$ &$\checkmark$& $\checkmark$\\ 
				\hline
				$A_8$&$\times$ & $\times$&$\times$ & $\times$ & $\checkmark$ & $\times$\\ 
				\hline
				$A_3/A_1$&$\checkmark$ & $\times$ &$\times$& $\checkmark$ &$\checkmark$& $\checkmark$\\ 
				\hline
				$A_3/A_5$&$\checkmark$ & $\checkmark^{(\times)}$ &$\times$ & $\checkmark$ &$\checkmark$& $\checkmark$\\ 
				\hline
				$A_6/A_8$&$\checkmark$ &$\checkmark^{(\times)}$ &$\times$ & $\checkmark$ &$\checkmark$& $\checkmark^{(\times)}$\\ 
				\hline
			\end{tabular}
			\end{center}
	\caption{\label{chart} \justifying The effectiveness of angular observables and their ratios in distinguishing the $g_R$ scenario from the SM and other NP scenarios. Results for $\mathcal{B}(B_c\rightarrow \tau \bar \nu_\tau)<30\%$ and $\mathcal{B}(B_c\rightarrow \tau \bar \nu_\tau)<10\%$ are shown.}
\end{table*}

\subsection{Impact of a tighter bound on $\mathcal{B}(B_c\rightarrow \tau \bar\nu_\tau)$}\label{sec:Br5}

\begin{figure}[t]
	\centering
	\includegraphics[width=\linewidth]{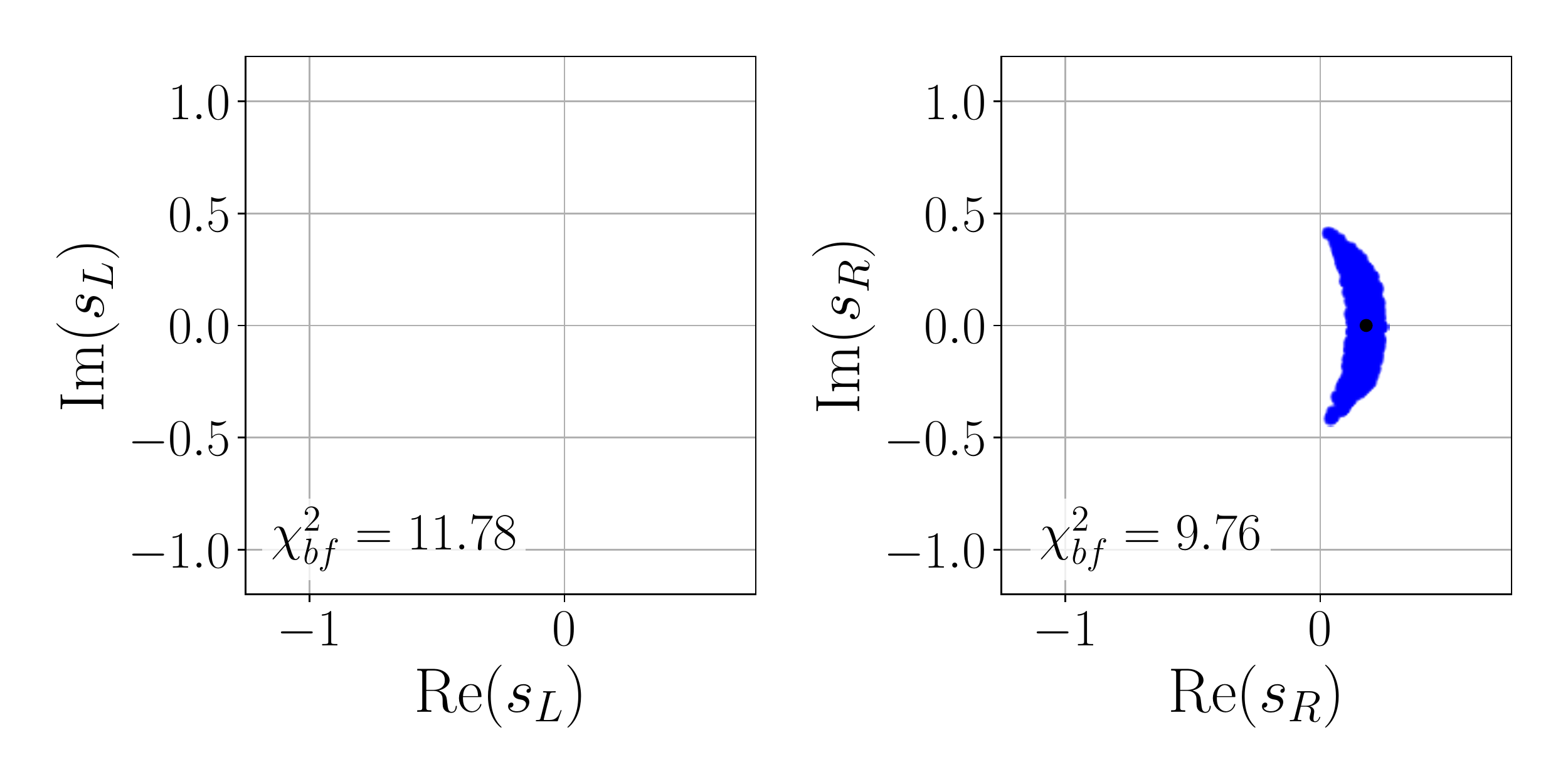}
	\caption{\justifying Allowed regions for $s_L$ and $s_R$ including the upper bound $\mathcal{B}(B_c\rightarrow \tau \bar \nu_\tau) < 10\%$. There are no $s_L$ values allowed at $2\sigma$. The blue region for $s_R$ corresponds to the parameter space allowed at $2\sigma$ (but not at $1.64\sigma$).}
	\label{fig: regionsBr10newChi} 
\end{figure}

In section \ref{obs}, we have analyzed the efficacy of the angular observables in distinguishing the $g_R$ scenario from other NP scenarios, considering the NP parameter spaces restricted  by $\mathcal{B}(B_c\rightarrow \tau \bar\nu_\tau)<30\%$. In this section we take a tighter bound $\mathcal{B}(B_c\rightarrow \tau \bar\nu_\tau)<10\%$  and discuss the effectiveness of the angular observables in distinguishing the effect of $g_R$. The allowed parameter spaces for $g_L$ and $g_R$ remain the same as in figure \ref{fig: regions}. The $s_L$ parameter space is totally excluded at $2\sigma$, while a small $s_R$ region around the best-fit value $s_{R,\,bf} = 0.18$ stays as shown in figure \ref{fig: regionsBr10newChi}. 


\begin{figure*}
	\centering
	\begin{subfigure}[b]{0.44\textwidth}
		\includegraphics[width=0.9\columnwidth]{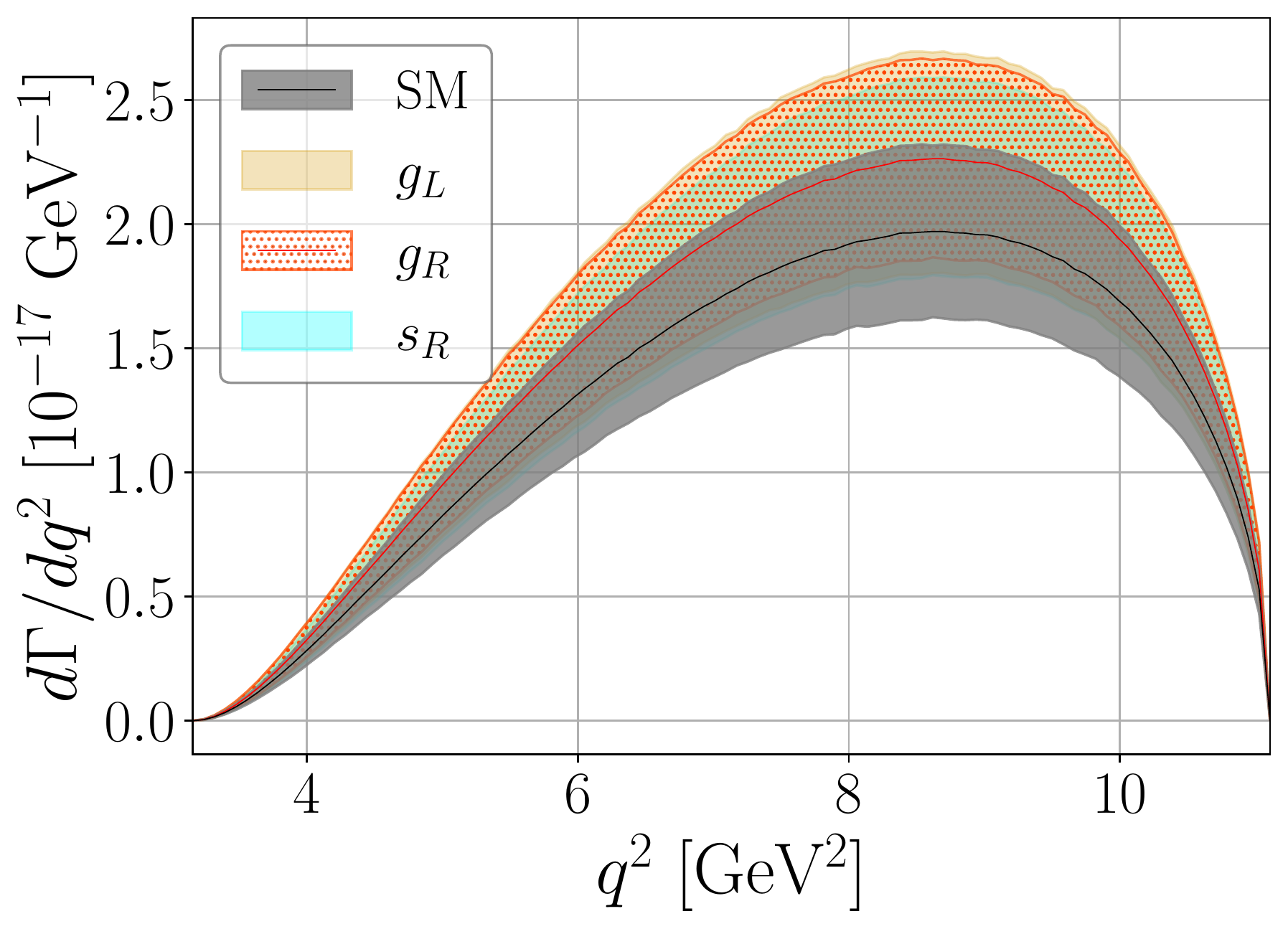}
	\end{subfigure}
	\hspace{0.2cm}
	\begin{subfigure}[b]{0.44\textwidth}
		\includegraphics[width=0.9\textwidth]{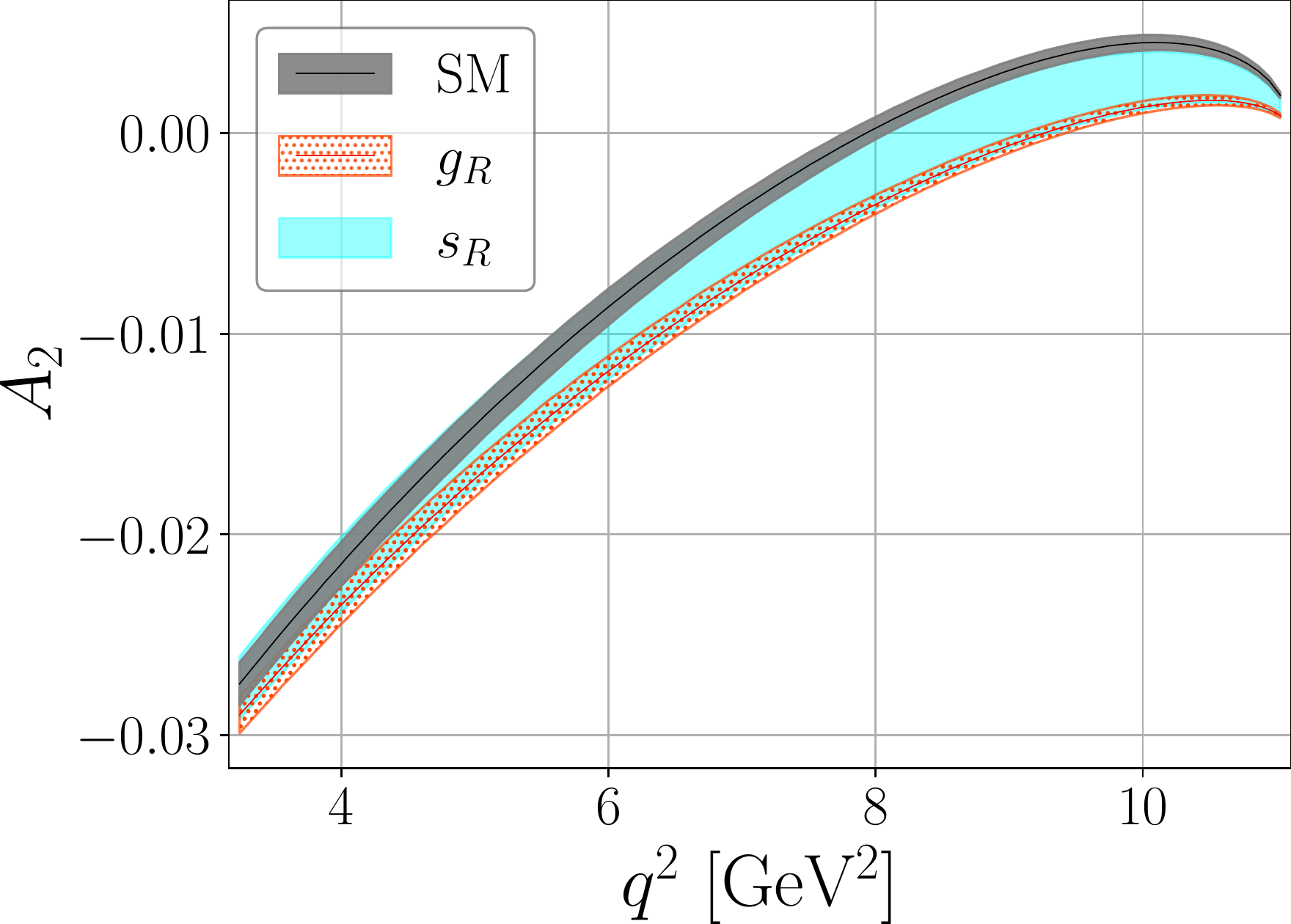}
	\end{subfigure}
	\begin{subfigure}[b]{0.44\textwidth}
		\includegraphics[width=0.9\textwidth]{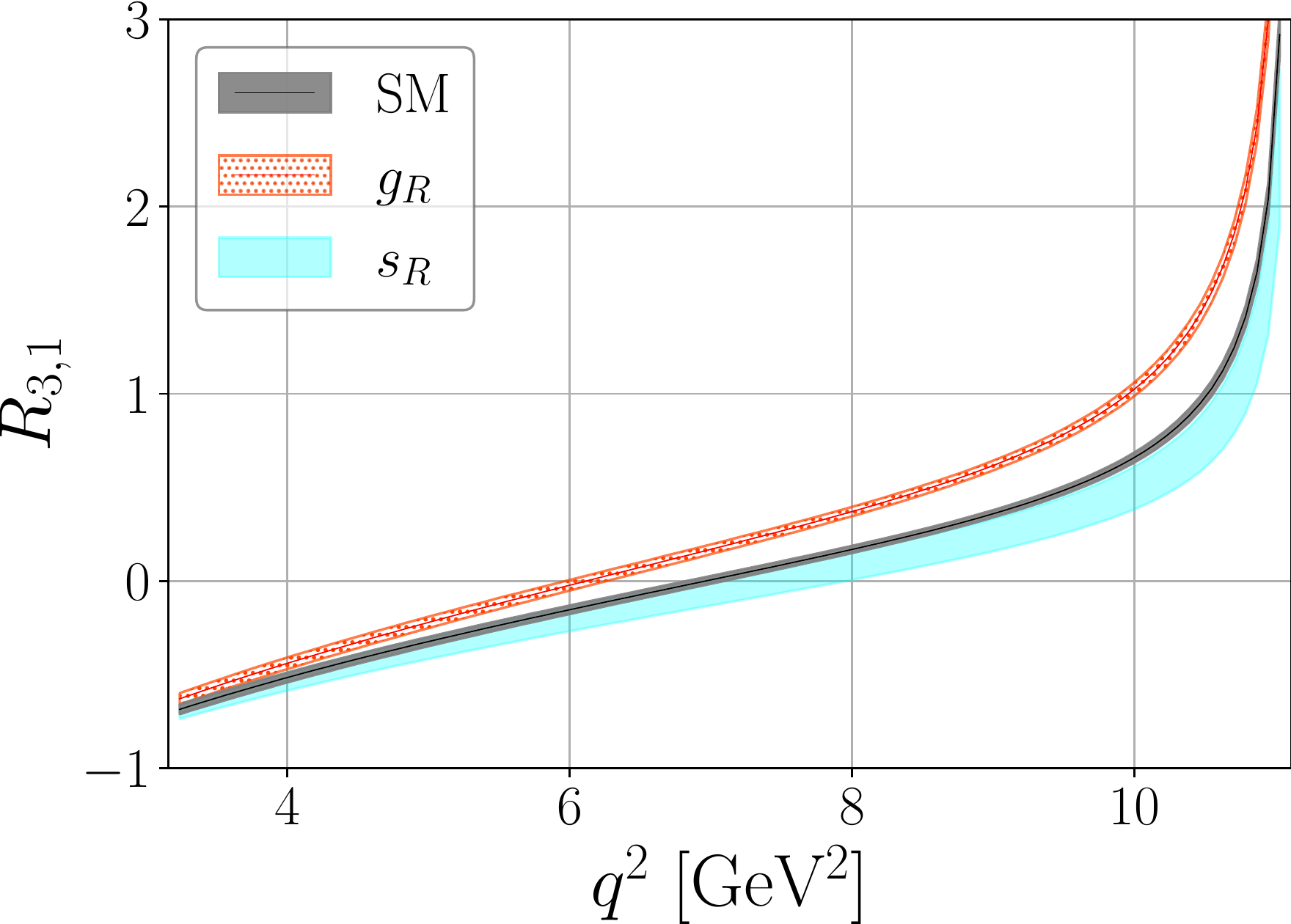}
	\end{subfigure}
	\hspace{0.2cm}
	\begin{subfigure}[b]{0.44\textwidth}
		\includegraphics[width=0.9\textwidth]{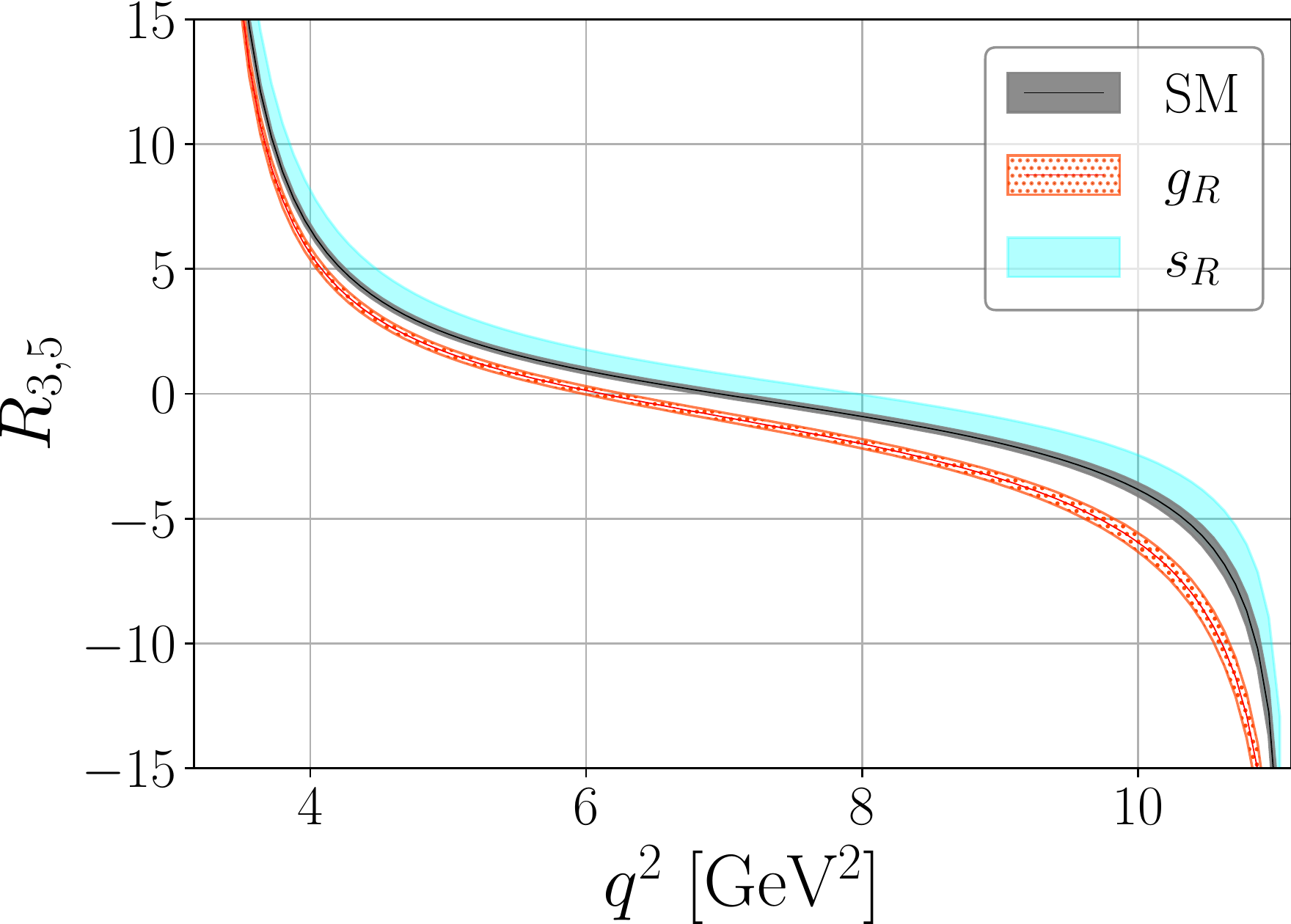}
	\end{subfigure}
	\begin{subfigure}[b]{0.44\textwidth}
		\includegraphics[width=0.9\textwidth]{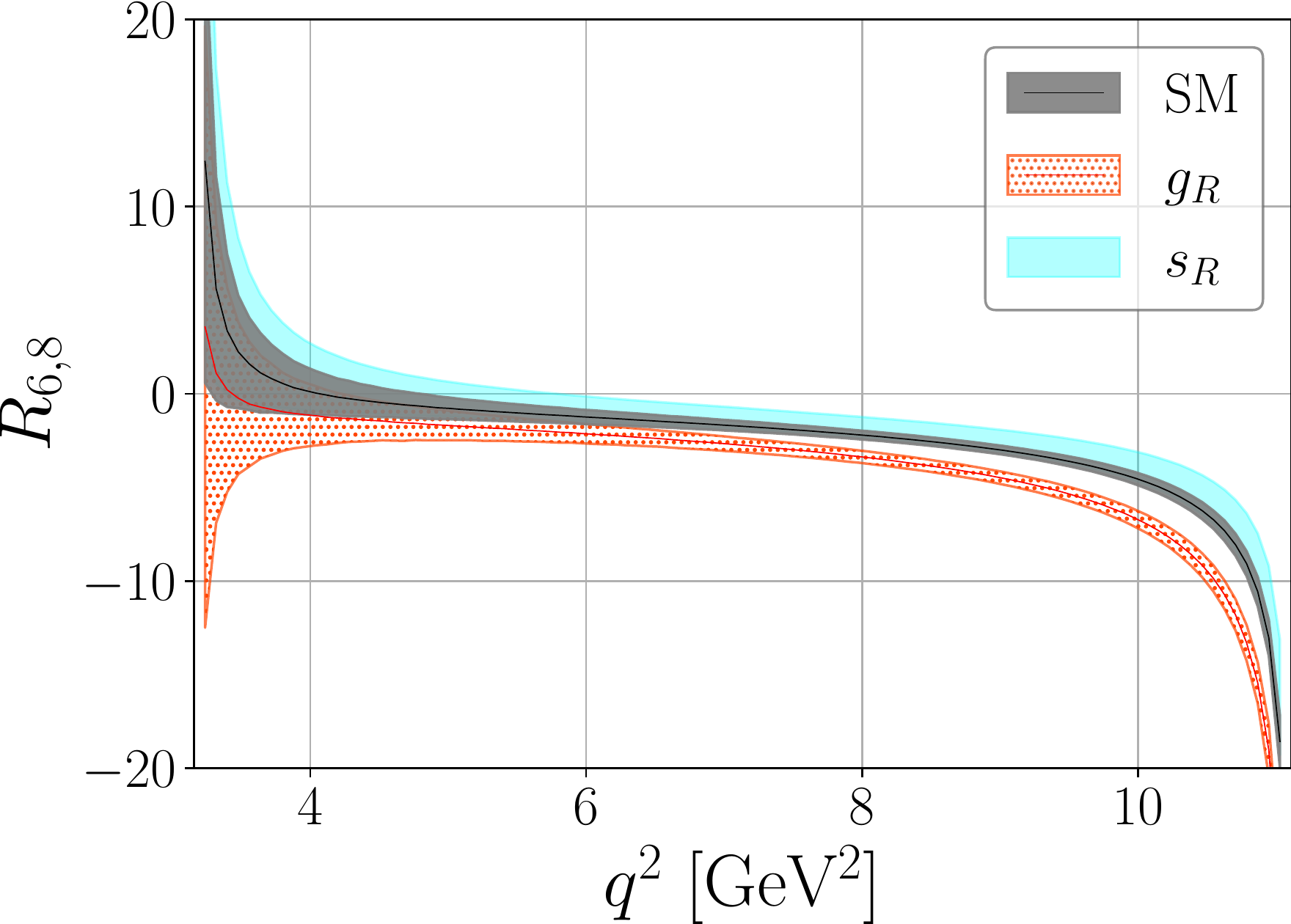}
	\end{subfigure}
	\hspace{0.2cm}
	\begin{subfigure}[b]{0.44\textwidth}
		\includegraphics[width=0.9\textwidth]{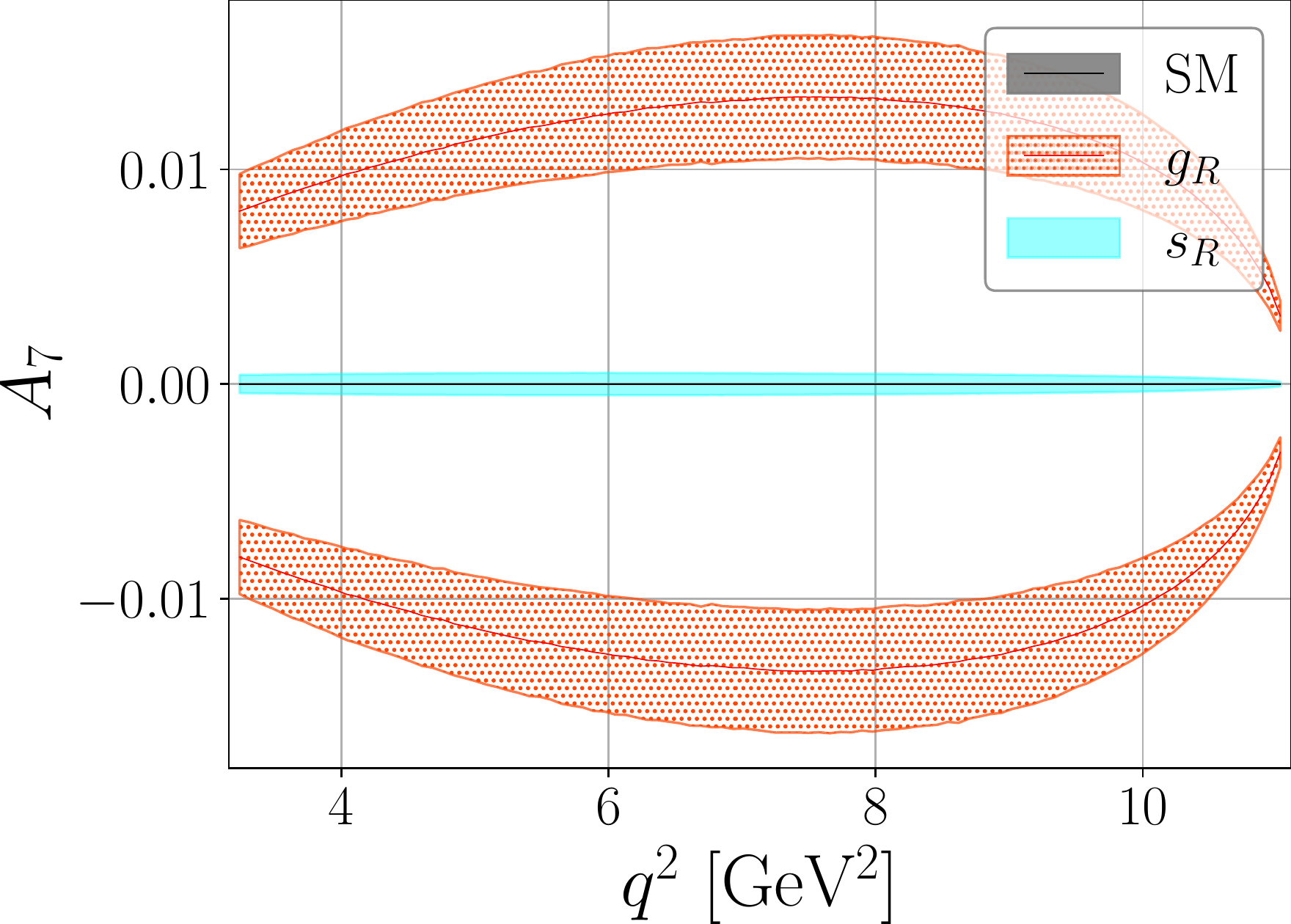}
	\end{subfigure}
	\caption{\justifying  The decay width $d\Gamma/dq^2$, the angular observables $A_2$, $A_7$ and the ratios $R_{3,1}$, $R_{3,5}$ and $R_{6,8}$ for $\mathcal{B}(B_c\rightarrow \tau \bar\nu_\tau)<10\%$ in SM and in the NP scenarios $g_L$, $g_R$ and $s_R$. The values of $g_L$ and $s_R$ are varied within their $2\sigma$-allowed range, while $g_R$ is kept fixed at its best-fit value as given in Table \ref{tab: bf2}. Note that there is no $s_L$ value allowed at $2\sigma$ with this upper bound on $\mathcal{B}(B_c\rightarrow \tau \bar\nu_\tau)$. For all observables except $d\Gamma/dq^2$, the results for $g_L$ are the same as those for SM.  For each scenario, $2\sigma$ uncertainties due to the hadronic form factors and the polarization asymmetry $\alpha_P$ have been included.}
	\label{fig: Br5all}
\end{figure*}	


The decay width $d\Gamma /dq^2$, the angular observables $A_2$, $A_7$ and the ratios $R_{3,1}$, $R_{3,5}$ and $R_{6,8}$ are shown in figure \ref{fig: Br5all}. With the tighter boundary condition $\mathcal{B}(B_c\rightarrow \tau \bar\nu_\tau)<10\%$, large regions of $s_L$ and $s_R$ parameter space are disallowed. As a result, a clear separation between the $g_R$ scenario and the $s_L$, $s_R$ scenarios is possible for multiple angular observables. 
\begin{itemize}
\item The angular observable $A_7$ and the ratios $R_{3,1}$ and $R_{3,5}$ can cleanly distinguish the effect of best-fit $g_R$ from the SM as well as from all the other NP scenarios considered here. 
\item The ratio $R_{6,8}$ can distinguish $g_R$ from the other scenarios at higher $q^2$ values.
\end{itemize}
In the right half of Table \ref{chart}, we summarize the results for all the angular observables and ratios with the condition $\mathcal{B}(B_c\rightarrow \tau \bar\nu_\tau)<10\%$. Note that we put a tick mark for all observables in the $s_L$ scenario because with the condition $\mathcal{B}(B_c\rightarrow \tau \bar \nu_\tau)<10\%$, there is no  $s_L$ value allowed at $2\sigma$. In Appendix \ref{app: bin-wise}, we show a comparison between the results obtained with the two upper bounds $\mathcal{B}(B_c\rightarrow \tau \bar\nu_\tau)<30\%$ and $\mathcal{B}(B_c\rightarrow \tau \bar\nu_\tau)<10\%$ for all points in the $2\sigma$-allowed region of $g_R$.

Note that this analysis has been performed in the idealized limit of a perfect detector and a large number of events. The systematic and statistical uncertainties in the experiment are not taken into account. However, the results above are expected to provide a good indication of which observables and which $q^2$ bins would be useful in distinguishing the $g_R$ scenario from the rest. Such a distinction will enable us to identify the presence of non-SMEFT contributions.

\section{Concluding remarks}\label{sec:conclusion}

The present measurements of production and decay channels of Higgs boson at the LHC are consistent with the SM Higgs mechanism of EW symmetry breaking. As a result, SMEFT is often taken to be the default EFT description above the electroweak scale. However, more general EFT descriptions such as HEFT, where the EW symmetry  $SU(2)_L\times U(1)_Y$ is realized non-linearly, are still possible. In this paper, we explore whether it is possible to identify HEFT signals that cannot be mimicked by operators allowed in SMEFT. We restrict ourselves to scalar and vector NP operators, neglecting any tensor contributions.

We focus on the flavor-physics sector, in particular on the six-dimensional flavor non-universal LEFT operator $O_V^{LR}\equiv(\bar{\tau}\gamma^\mu P_L\nu_\tau)(\bar{c}\gamma_\mu P_R b)$  which is relevant for $b\rightarrow c \tau \nu_\tau$ processes. This operator can arise from a six-dimensional HEFT operator. However, it appears in SMEFT only at dimension-8 and hence its contribution would be suppressed by an extra factor of $v^2/\Lambda^2$. Therefore, a large contribution from this operator would prefer HEFT over SMEFT as the correct BSM description.

The principle of EFTs is that any operator that is not explicitly forbidden by a symmetry must be taken into account, which is the case for $O_V^{LR}$ in a beyond-SMEFT scenario. Therefore, the consideration of this operator becomes mandatory even in the absence of a UV-complete model. There are also specific models, like the nonstandard-Higgs model with a strongly-coupled scalar~\cite{Cata:2015lta} and the one with a $W'$ that couples to right-handed quarks and left-handed leptons~\cite{Mu:2019bin} that give rise to the $O_V^{LR}$ operator.

The decay  $\Lambda_b\rightarrow\Lambda_c(\Lambda\pi)\tau\bar{\nu}_\tau$, when fully reconstructed, can be written in terms of an angular distribution in three angles. The coefficients of these angular terms, $A_0$, ..., $A_9$, can act as angular observables where information about NP is encoded. We calculate this angular distribution in terms of helicity amplitudes and compare the values of the above angular observables in scenarios with (i) only SM, (ii) NP with non-zero coefficient $g_R$ of $O_V^{LR}$, (iii) NP with non-zero coefficient $g_L$ of $O_V^{LL}\equiv (\bar{\tau}\gamma^\mu P_L\nu_\tau)(\bar{c}\gamma_\mu P_Lb)$, (iv)  NP with new scalar and pseudo-scalar operators parameterized by non-zero values of $s_L$ and $s_R$. The identification of the $g_R$ scenario would indicate the need for a HEFT description, i.e., for going beyond SMEFT. 

We find that the angular observables $A_2$, $A_7$, and the ratios $R_{3,1}\equiv A_3/A_1$,  $R_{3,5}\equiv A_3/A_5$ and $R_{6,8}\equiv A_6/A_8$ are in principle capable of distinguishing the effects of $g_R$ from those of the SM and $g_L$ scenarios. These can thus indicate the presence of BSM physics. It is observed that their effectiveness strongly depends on the constraints on the branching ratio  $\mathcal{B}(B_c\rightarrow \tau \bar \nu_\tau)$. For a weaker bound $\mathcal{B}(B_c\rightarrow \tau \bar \nu_\tau)<30\%$, only the angular observable $A_7$ can cleanly distinguish the effects of $g_R$ from those of the $s_L$ and $s_R$ scenarios as well. However, if a tighter bound $\mathcal{B}(B_c\rightarrow \tau \bar \nu_\tau)<10\%$ can be imposed, we find that along with $A_7$, the ratios $R_{3,1}$, $R_{3,5}$ and $R_{6,8}$ could also indicate distinct effects for $g_R$ in comparison to SM and other NP scenarios. For all the cases, this distinction would be facilitated in higher-$q^2$ bins and for larger values of $|\textrm{Im}(g_R)|$. Clearly, constraining the branching ratio of $B_c\rightarrow \tau \bar \nu_\tau$ would be extremely crucial in identifying physics beyond SMEFT. Along with more data for this process, a better understanding of the $p_T$-dependence of the fragmentation function in $B_c$ decay will be needed to put a tighter bound on this branching ratio.

The decay of $\Lambda_b\rightarrow\Lambda_c(\Lambda\pi)\tau\bar{\nu}_\tau$ has been recently observed at LHCb. Future runs of High-Luminosity LHC is expected to provide more data in this channel.  With enough data, the angular distribution in this channel could allow us to identify the the presence of the NP operator $O_V^{LR}$, which will indicate that SMEFT is not a sufficient EFT description of BSM physics beyond the EW scale. This, in turn, would help us probe the nature of Higgs and the mode of realization of $SU(2)_L\times U(1)_Y$ symmetry above the EW scale.


\begin{acknowledgements}
We would like to thank Rick S. Gupta, Soumen Halder, Gagan B. Mohanty, Arnab Roy and Tuhin S. Roy for useful discussions. This work is supported by the Department of Atomic Energy, Government of India, under Project Identification Number RTI 4002.  We acknowledge the use of computational facilities of the Department of Theoretical Physics at Tata Institute of Fundamental Research, Mumbai. We would also like to thank Ajay Salve and  Kapil Ghadiali for technical assistance. 
\end{acknowledgements}


\appendix
\section{Hadronic amplitudes}\label{app: hadronic}

  As mentioned in section \ref{sec:angdist}, eq.\,(\ref{hadronicdef}), the hadronic amplitudes are nonzero only for the combinations
  \begin{align} 
  H^{VA}_{\lambda_2,\lambda_3}& =H^V_{\lambda_2,\lambda_3}-H^A_{\lambda_2,\lambda_3}~,\\
  H^{SP}_{\lambda_2,\lambda_3}&=H^S_{\lambda_2,\lambda_3}+H^P_{\lambda_2,\lambda_3}~.
  \end{align}
 In this section, we provide explicit expressions for these combinations for all $\lambda_2$ - $\lambda_3$ pairs, in terms of the hadronic form factors $F_+,~F_\perp,~F_0,~G_+,~G_\perp$ and $G_0$~\cite{Datta:2017aue}.
  
  The hadronic amplitudes involving vector currents are shown in the following.
\begin{align}
	H^{VA}_{\frac{1}{2},0}&=F_+ (g_L+g_R+1) (m_{\Lambda_b}+m_{\Lambda_c}) \sqrt{Q_-/q^2}\nonumber\\
	&~-G_+ (g_L-g_R+1) (m_{\Lambda_b}-m_{\Lambda_c}) \sqrt{Q_+/q^2}~,\nonumber\\
	H^{VA}_{\frac{1}{2},1}&=G_\perp (g_L-g_R+1) \sqrt{2 Q_+}\nonumber\\
	&~-F_\perp (g_L+g_R+1) \sqrt{2 Q_-}~,\nonumber\\
	H^{VA}_{\frac{1}{2},t}&=F_0 (g_L+g_R+1) (m_{\Lambda_b}-m_{\Lambda_c}) \sqrt{Q_+/q^2}\nonumber\\
	&~-G_0 (g_L-g_R+1) (m_{\Lambda_b}+m_{\Lambda_c}) \sqrt{Q_-/q^2}~,\nonumber\\
	H^{VA}_{-\frac{1}{2},0}&=F_+ (g_L+g_R+1) (m_{\Lambda_b}+m_{\Lambda_c}) \sqrt{Q_-/q^2}\nonumber\\
	&~+G_+ (g_L-g_R+1) (m_{\Lambda_b}-m_{\Lambda_c}) \sqrt{Q_+/q^2}~,\nonumber\\
	H^{VA}_{-\frac{1}{2},-1}&=-F_\perp (g_L+g_R+1) \sqrt{2 Q_-}\nonumber\\
	&~-G_\perp (g_L-g_R+1) \sqrt{2 Q_+}~,\nonumber\\
	H^{VA}_{-\frac{1}{2},t}&=F_0\,(g_L+g_R+1) (m_{\Lambda_b}-m_{\Lambda_c}) \sqrt{Q_+/q^2}\nonumber\\
	&~+G_0 (g_L-g_R+1) (m_{\Lambda_b}+m_{\Lambda_c}) \sqrt{Q_-/q^2}~.
\end{align}
 Here $Q_+ = (m_{\Lambda_b}+m_{\Lambda_c})^2-q^2$ and $Q_- = (m_{\Lambda_b}-m_{\Lambda_c})^2-q^2$, with $q^2$ being the total invariant mass squared of the leptons $\tau$ and $\bar\nu_\tau$. The suffix $t$ indicates the contribution from $s_3=0$. 
The values of hadronic form factors $F_+,~F_\perp,~F_0,~G_+,~G_\perp$ and $G_0$ are as obtained in~\cite{Detmold:2015aaa} using the ``nominal fit". 

The hadronic amplitudes involving scalar currents are 
\begin{align}
	H^{SP}_{\frac{1}{2},t}&=\frac{F_0\; g_S (m_{\Lambda_b}-m_{\Lambda_c}) \sqrt{Q_+}}{m_b-m_c}\nonumber\\
	&~-\frac{G_0\; g_P (m_{\Lambda_b}+m_{\Lambda_c}) \sqrt{Q_-}}{m_b+m_c}~,\nonumber
\end{align}
\begin{align}
	H^{SP}_{-\frac{1}{2},t}&=\frac{F_0\; g_S (m_{\Lambda_b}-m_{\Lambda_c}) \sqrt{Q_+}}{m_b-m_c}\nonumber\\
	&~+\frac{G_0\; g_P (m_{\Lambda_b}+m_{\Lambda_c}) \sqrt{Q_-}}{m_b+m_c}~.\label{HVA}
\end{align}

\section{Extracting angular observables}\label{app:moments}

The coefficients of the angular distribution in eq.\,(\ref{angdist}) can be extracted by a fit to the $10$ coefficients. Alternatively, they can be extracted by using the orthogonality between different angular functions, using the method of angular moments~\cite{Dighe:1998vk}.  We can write the angular distribution in the following form:
\begin{align}
	\frac{1}{d\Gamma/dq^2}\frac{d\Gamma}{dq^2 d\cos\theta_c \,d\cos\theta_l d\chi}&=\sum_i A_i g_i(\cos\theta_c,\cos\theta_l,\chi)~.
\end{align} 
Here $A_i$ are the coefficients of the angular distribution in eq.\,(\ref{angdist}) and $g_i(\cos\theta_c,\cos\theta_l,\chi)$ are the associated angular functions. We find a ``weighting function" $w_i(\cos\theta_c,\cos\theta_l,\chi)$ for each of the $g_i$ such that 

\begin{align}
	&\int_{-1}^{1}d\cos\theta_l\int_{-1}^{1}d\cos\theta_c\int_{0}^{2\pi}d\chi\nonumber\\
	&~ g_i(\cos\theta_c,\cos\theta_l,\chi)\, w_j(\cos\theta_c,\cos\theta_l,\chi) = \delta_{ij}~,
\end{align}
where $\delta_{ij}$ is the Kronecker delta function. The value of $A_i$ in a particular $q^2$ bin can then be determined as

\begin{equation}
	A_i (\textrm{bin})= \frac{1}{N_{\rm{bin}}}\sum_{\rm{events\in \rm{bin}}}w_i(\cos\theta_c,\cos\theta_l,\chi)~.\label{angularmoment}
\end{equation}
The weighting functions $w_i(\cos\theta_c,\cos\theta_l,\chi)$ are given in Table~\ref{angularmomenttable}. Note that these weighting functions are not unique and may not be experimentally optimal.

\begin{table}[h!]
	\begin{center}
		\begin{tabular}{|c|c|}
			\hline
			Coefficient& Weighting Function ($w_i$)\\
			\hline&\\[-1em]
			$A_0$ & $ \frac{1}{8\pi}+\frac{5}{32\pi}(1-3\cos^2\theta_l)$\\[5pt]
			\hline&\\[-1em]
			$A_1$ & $ \left(\frac{3}{8\pi}+\frac{15}{32\pi}(1-3\cos^2\theta_l)\right)\cos\theta_c$ \\[5pt]
			\hline&\\[-1em]
			$A_2$ & $\frac{3}{8\pi}\cos\theta_l$\\[5pt]
			\hline&\\[-1em]
			$A_3$ & $\frac{9}{8\pi}\cos\theta_c \cos\theta_l$\\[5pt]
			\hline&\\[-1em]
			$A_4$ & $-\frac{15}{32\pi}(1-3\cos^2\theta_l)$\\[5pt]
			\hline&\\[-1em]
			$A_5$ & $ -\frac{45}{32\pi}(1-3\cos^2\theta_l)\cos\theta_c$\\[5pt]
			\hline&\\[-1em]
			$A_6$ & $\frac{4}{\pi^3}\cos\chi$\\[5pt]
			\hline&\\[-1em]
			$A_7$ & $\frac{4}{\pi^3}\sin\chi$\\[5pt]
			\hline&\\[-1em]
			$A_8$ & $\frac{16}{\pi^3}\cos\chi\cos\theta_l$\\[5pt]
			\hline&\\[-1em]
			$A_9$ & $\frac{16}{\pi^3}\sin\chi\cos\theta_l$\\[5pt]
			\hline
		\end{tabular}
	\end{center}
	\caption{\justifying Weighting functions to extract the coefficients of angular terms from the angular distribution.}
	\label{angularmomenttable}
\end{table}


\section{Mapping the angular distribution onto earlier literature}\label{app:mapping}

We compare our expressions of the angular distribution with \cite{Boer:2019zmp}, where a discussion about the  angular distributions in earlier literature is provided. In~\cite{Boer:2019zmp}, the angular distribution of $\Lambda_b\rightarrow\Lambda_c(\Lambda\pi)\tau\bar{\nu_\tau}$ decay is given as
\begin{align}
	&K(q^2,\cos\theta_l^\prime,\cos\theta_c^\prime,\chi^\prime)\nonumber\\
	&\equiv\frac{8\pi}{3}\frac{1}{d\Gamma/d q^2}\frac{d^4\Gamma}{dq^2 d\ cos\theta_l^\prime d\cos\theta_c^\prime d \chi^\prime}\nonumber\\
	&=K_{1ss}\sin^2\theta_l^\prime+K_{1cc}\cos^2\theta_l^\prime+K_{1c}\cos\theta_l^\prime\nonumber\\
	&+(K_{2ss}\sin^2\theta_l^\prime++K_{2cc}\cos^2\theta_l^\prime+K_{2c}\cos\theta_l^\prime)\cos\theta_c^\prime\nonumber\\
	&+(K_{3sc}\sin\theta_l^\prime\cos\theta_l^\prime+K_{3s}\sin\theta_l^\prime)\sin\theta_c^\prime\sin\chi^\prime\nonumber\\
	&+(K_{4sc}\sin\theta_l^\prime\cos\theta_l^\prime+K_{4s}\sin\theta_l^\prime)\sin\theta_c^\prime\cos\chi^\prime~.\label{angdistboer}
\end{align}
The angles in the above equation are related to the angles defined in section~\ref{sec:angdist} as
\begin{equation}
	\theta_c^\prime\longrightarrow\theta_c~,~~~~~\theta_l^\prime\longrightarrow\pi-\theta_l~~~\textrm{and}~~\chi^\prime\longrightarrow\chi~.
\end{equation}
The angular coefficients in eq.\,(\ref{angdist}) can be mapped onto the coefficients in eq.\,(\ref{angdistboer}) as shown below.
\begin{align}
	K_{1ss}\rightarrow (8\pi/3)\,A_0,\quad & K_{1cc}\rightarrow (8\pi/3)(A_0+A_4), \\
	\,K_{1c}\rightarrow -(8\pi/3)A_2, \quad& K_{2ss}\rightarrow (8\pi/3)\,A_1,\\
	K_{2c}\rightarrow -(8\pi/3)\,A_3,\quad	& K_{2cc}\rightarrow (8\pi/3)\,(A_1+A_5),\\
	K_{3sc}\rightarrow (8\pi/3)\,A_9, \quad&  K_{3s}\rightarrow (8\pi/3)\,A_7, \\
	K_{4sc}\rightarrow -(8\pi/3)\,A_8,\quad & K_{4s}\rightarrow (8\pi/3)\,A_6~.
\end{align}
We have chosen the convention for the angles following~\cite{kutsckhe1996angular}, as it provides a general procedure to calculate angular distribution using helicity amplitudes for many such decays. This convention also matches with~\cite{Gutsche:2015rrt, Mu:2019bin}. The form of our angular terms is similar to that in~\cite{Becirevic:2022bev}, where the angular distribution is given in term of helicities of $\Lambda$ and $\tau$.		

\section{Beyond-SMEFT effects in different $q^2$ bins over the $g_R$ parameter space}\label{app: bin-wise}

\begin{figure*}
	\centering
	\includegraphics[width=.96\textwidth]{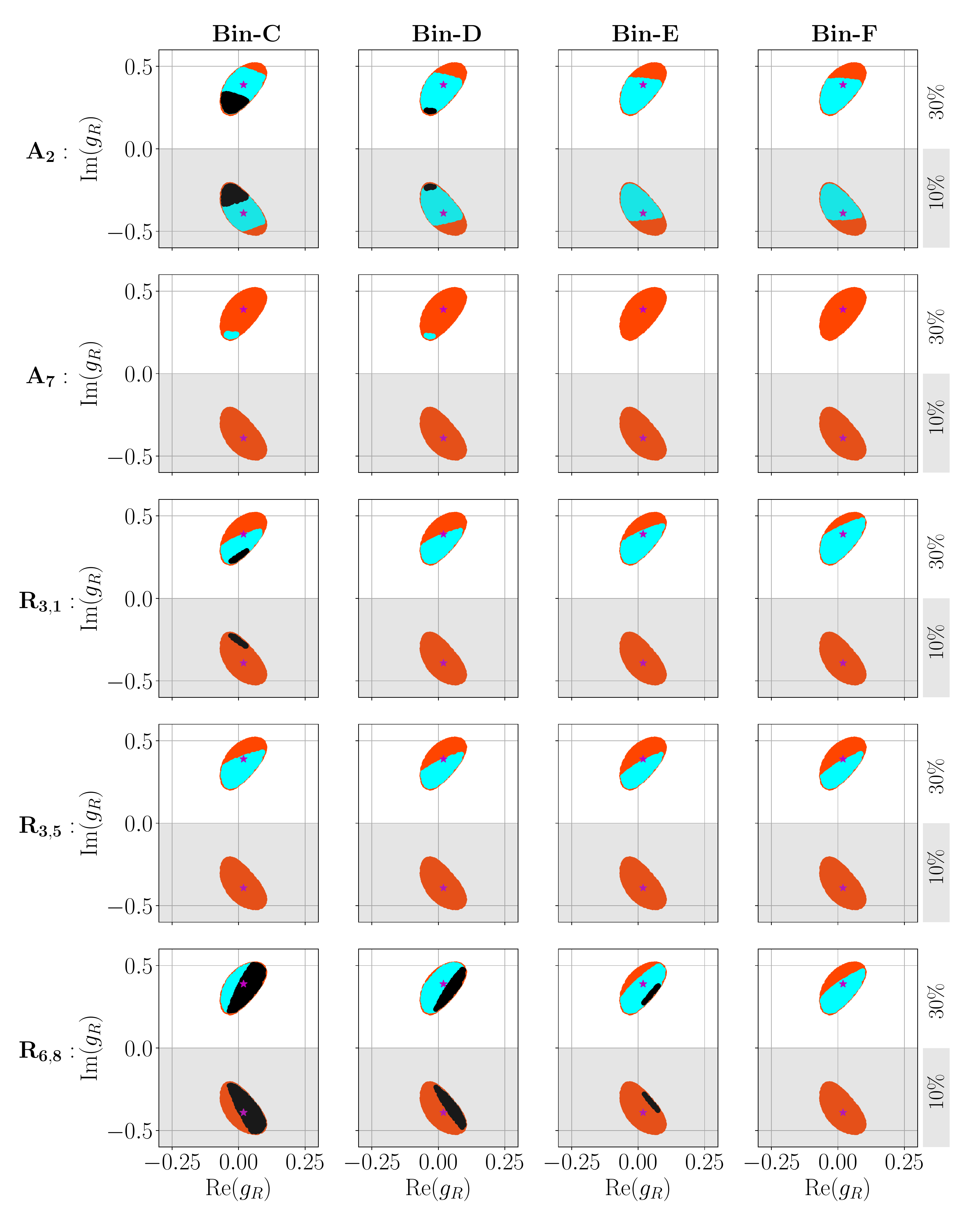}
	\caption{\justifying  The $2\sigma$-allowed parameter space of $g_R$, where the $g_R$ scenario can be distinguished from all the other scenarios \{SM, $g_L$, $s_L$, $s_R$\} (orange), from only SM and $g_L$ (cyan), and from neither (black). The best-fit value of $g_R$ is denoted by a star. The upper half of each panel (unshaded) corresponds to results for  $\mathcal{B}(B_c\rightarrow \tau \bar \nu_\tau)<30\%$, while the lower half (shaded) corresponds to  $\mathcal{B}(B_c\rightarrow \tau \bar \nu_\tau)<10\%$.}
	\label{gRprobed}
\end{figure*}

In section \ref{sec:results}, we presented our results with the best-fit value of $g_R$ and discussed the possibility of distinguishing it from SM and from the NP scenarios $g_L$, $s_L$ and $s_R$. In this section, we vary $g_R$ over its $2\sigma$-allowed range, and indicate those regions  for which the distinction from the other scenarios would be theoretically clean in a given $q^2$ bin. By this criteria, we mean that the theoretical prediction for the given observable in the given $q^2$ bin has no overlap between the $g_R$ scenario and the other scenarios, even taking into account the theoretical uncertainties. Note that this approach as presented in figure \ref{gRprobed} is a crude way of recording the separation of bands in figures \ref{A2fig}, \ref{A135}, \ref{A678} and \ref{fig: Br5all}, in a binary manner (non-zero overlap or no overlap). It does not necessarily correspond to experimental feasibility of identifying $g_R$.


The analysis of $\Lambda_b\rightarrow\Lambda_c(\rightarrow\Lambda\pi)\tau\bar\nu_\tau$ in the LHCb experiment~\cite{LHCb:2022piu} is carried out by dividing the data in six $q^2$ bins, in the range $0~\textrm{to}~(m_{\Lambda_b}-m_{\Lambda_c})^2$. The bins (A, B, C, D, E, F) correspond to the $q^2$ ranges ($0-1.83$, $1.83-3.67$, $3.67-5.5$, $5.50-7.33$, $7.33-9.17$, $9.17-11.13$) $\textrm{GeV}^2$. The first bin cannot have any signal events since $m_\tau^2=3.16$ GeV$^2$, while bin B also does not give any significant number of events~\cite{LHCb:2022piu}. Therefore we focus only on the four bins C, D, E and F. Note that the angular analysis performed in the actual experimental analysis may not follow the same $q^2$ binning scheme as presented in \cite{LHCb:2022piu}. Our choice of bins is only for the purpose of illustration. Presenting the results in different $q^2$ bins highlights the fact that the separation of the $g_R$ scenario from the other scenarios may not be possible in all the bins but only in some subsets of them. Note that when clear separation happens in any of the bins, it indicates a signal of physics beyond SMEFT.

For every point in the $2\sigma$-allowed region of $g_R$ in our analysis, we check if the effects of $g_R$ can be distinguishable  (i) from SM and $g_L$,  (ii) from other NP scenarios, $s_L$ and $s_R$, in each of these four bins.  In figure~\ref{gRprobed}, we represent the $g_R$ values that can be distinguishable from both (i) and (ii) by orange color. Those $g_R$ which can be distinguishable from (i) but not from (ii) are denoted by cyan color, while those values that can be distinguishable neither from (i) nor from (ii) are denoted by black.  Here, by being distinguishable we mean that the predicted values of the observable in that bin for that particular $g_R$ value do not overlap with the predicted values for the scenarios (i) or (ii). We take into account the $2\sigma$ uncertainties due to the hadronic form factors and the polarization asymmetry $\alpha_P$, as earlier. 

Note that the constraints on $g_R$ used in our fit in section \ref{sec:results} are independent of the sign of $\rm{Im}(g_R)$. Therefore, the allowed region of $g_R$ is symmetric with respect to the real axis. We use this fact to show the comparison between the two scenarios $\mathcal{B}(B_c\rightarrow \tau \bar \nu_\tau)<30\%$ and $\mathcal{B}(B_c\rightarrow \tau \bar \nu_\tau)<10\%$ in a compact manner.  We use the upper half of the region (unshaded) to show the constraints when $\mathcal{B}(B_c\rightarrow \tau \bar \nu_\tau)<30\%$ and the lower half (shaded) to show the constraints when $\mathcal{B}(B_c\rightarrow \tau \bar \nu_\tau)<10\%$. The presence of the corresponding regions reflected about the $\textrm{Re}(g_R)$ axis is implicit.

Based on the discussion in section~\ref{sec:results}, we focus on the angular observables $A_2$, $A_7$,  and the ratios $R_{3,1}$, $R_{3,5}$ and $R_{6,8}$, as they are expected to be more suitable in distinguishing $g_R$ from the other scenarios. We note the following: 
\begin{itemize}
	\item  Using $A_7$, it may be possible to distinguish $g_R$ from $SM$ as well as from all the other NP scenarios in most of the allowed  $2\sigma$ range of $g_R$.
	\item With the loose constraint $\mathcal{B}(B_c\rightarrow \tau \bar \nu_\tau)<30\%$, we find that the ratios $R_{3,1}$, $R_{3,5}$ and $R_{6,8}$ and the angular observable $A_2$ may be able to distinguish a large region of $g_R$ from SM and $g_L$, but would be unable to distinguish it from $s_L$ or $s_R$. Only a small $g_R$ parameter space can be distinguishable from the effects of $s_L$ and $s_R$ in higher $q^2$-bins. 
	\item On the other hand, the tighter constraint $\mathcal{B}(B_c\rightarrow \tau \bar \nu_\tau)<10\%$ allows the ratios $R_{3,1}$ and $R_{3,5}$ to distinguish large regions in the $g_R$ parameter space from all the other scenarios.
 
	\item  The power of distinction is typically higher for larger values of $|\textrm{Im}(g_R)|$.
	\item For all the five observables, the power of distinguishing $g_R$ from the other scenarios appears to be higher at larger $q^2$. 
\end{itemize}





\end{document}